\documentclass[11pt,a4paper]{article}
\usepackage[sfdefault]{overlock}
\usepackage[left=2cm,right=2cm,top=2cm,bottom=2cm]{geometry}
\usepackage[utf8]{inputenc}
\usepackage[english]{babel}
\usepackage[T1]{fontenc}
\usepackage{amsmath}
\usepackage{amsthm}
\usepackage{amsfonts}
\usepackage{dsfont}
\usepackage{mathrsfs}
\usepackage{amssymb}
\usepackage{graphicx}
\usepackage[usenames,dvipsnames,svgnames,table]{xcolor}
\usepackage{stmaryrd}
\usepackage{appendix}
\usepackage{enumerate}
\usepackage{enumitem}
\usepackage{algpseudocode}
\usepackage{algorithm}
\usepackage{thmtools}
\usepackage{nameref,cleveref}
\usepackage{xspace}
\usepackage[affil-it]{authblk}
\usepackage[square, authoryear]{natbib}
\bibpunct{[}{]}{ ;}{a}{}{;}

\newtheorem{theo}{Theorem}
\newtheorem{cor}[theo]{Corollary}
\newtheorem{prop}[theo]{Proposition}
\newtheorem{assp}{Assumption}
\newtheorem{lem}[theo]{Lemma}

\Crefname{theo}{Theorem}{Theorems}
\crefname{theo}{theorem}{theorems}
\Crefname{assp}{Assumption}{Assumptions}
\crefname{assp}{assumption}{assumptions}
\Crefname{prop}{Proposition}{Propositions}
\crefname{prop}{proposition}{propositions}
\Crefname{lem}{Lemma}{Lemmas}
\crefname{lem}{lemma}{lemmas}
\Crefname{rmk}{Remark}{Remarks}
\crefname{rmk}{remark}{remarks}
\newcommand{\Rodeo}{\textsc{Rodeo}\xspace}
\newcommand{\CDRodeo}{\textsc{CDRodeo}\xspace}
\newcommand{\R}{\mathbb{R}}
\newcommand{\nX}{n_X}

\newcommand{\PP}[1]{\mathbb{P}\left(#1\right)}
\newcommand{\E}[1]{\mathbb{E}\left[ #1 \right]}
\newcommand{\V}[1]{\text{Var}\left(#1\right)}
\newcommand{\ind}{\mathds{1}}

\newcommand{\rel}{\mathcal{R}}
\newcommand{\preA}{\tau_{n}}
\newcommand{\Ctau}{\text{C}_{\tau}}
\newcommand{\postA}{T_{n}}
\newcommand{\CT}{\text{C}_{T}}

\newcommand{\seuil}[1]{\lambda_{#1}}
\newcommand{\Cl}{\text{C}_{\lambda}}
\newcommand{\Un}{\ \mathcal{U}_n(x)}
\newcommand{\Up}{\ \mathcal{U}'_n(x)}

\newcommand{\Ux}[1]{\ \mathcal{U}_{#1}(x)}
\newcommand{\UUn}{\ \mathcal{U}_n(w)}

\newcommand{\supp}{\text{supp}}
\newcommand{\HGP}{\mathcal{H}_{\text{hp}}}

\newcommand{\f}[1]{\hat{f}_{#1}(w)}
\newcommand{\fX}{\tilde{f}_X}
\newcommand{\An}{\widetilde{A}_n}
\newcommand{\fXK}{\tilde{f}_{X}^K}
\newcommand{\fXKi}[1]{\tilde{f}_{X #1}^K}

\newcommand{\hX}{h_X}
\newcommand{\KX}{K_X}
\newcommand{\X}{\widetilde{X}}
\newcommand{\dX}{\widetilde{\delta}_{X}}

\newcommand{\fbar}[1]{\bar{f}_{#1}(w)}
\newcommand{\Zbar}[1]{\bar{Z}_{#1}}

\newcommand{\gZ}{\gamma_{Z,n}}
\newcommand{\CgfA}{\text{C}_{\gamma \vert f\vert}}
\newcommand{\CgZA}{\text{C}_{\gamma \vert \bar{Z} \vert}}
\newcommand{\CEZ}{\text{C}_{E\bar{Z},j}}

\newcommand{\CEZA}{\text{C}_{E\vert \bar{Z} \vert}}
\newcommand{\condZbar}{\text{cond}_{\bar{Z}}}
\newcommand{\CMZbar}{\text{C}_{\text{M}\bar{Z}}}
\newcommand{\MZbar}[1]{\text{M}_{\bar{Z},#1}}
\newcommand{\Mbar}[1]{\overline{\text{M}}_{#1}}
\newcommand{\CMbar}{\text{C}_{\bar{\text{M}}}}
\newcommand{\vZbar}[1]{\text{v}_{\bar{Z},#1}}
\newcommand{\CvZbar}{\text{C}_{\text{v}\bar{Z}}}
\newcommand{\DZ}[1]{\Delta_{Z,#1}}

\newcommand{\CMDZ}{\text{C}_{\text{M}\Delta Z}}
\newcommand{\MD}[1]{\text{M}_{\Delta}}
\newcommand{\CMD}{\text{C}_{\text{M}\Delta}}
\newcommand{\BZA}[1]{\mathcal{B}_{\vert \bar{Z}\vert, #1}}
\newcommand{\BZ}[1]{\mathcal{B}_{\bar{Z}, #1}}
\newcommand{\B}[1]{\overline{\mathcal{B}}_{ #1}}
\newcommand{\BA}[1]{\mathcal{B}_{\vert\bar{f}\vert#1}}
\newcommand{\EZ}{\mathcal{E}_{Z}}
\newcommand{\Ef}{\mathcal{E}_{f}}

\newcommand{\Cbiasbar}{\text{C}_{\overline{\text{bias}}}}
\newcommand{\CbiasX}{\text{C}_{\text{bias}_X}}
\newcommand{\CEbar}{\text{C}_{\bar{\text{E}}}}
\newcommand{\Vbar}[1]{\overline{\text{v}}_{#1}}

\newcommand{\Cs}{\text{C}_{\sigma}}

\begin{document}
\author{Minh-Lien Jeanne Nguyen 
}
\affil{Laboratoire de Mathématiques d’Orsay\\
Univ. Paris-Sud, CNRS, Université Paris-Saclay, 91405 Orsay, France}
\title{\bf  Nonparametric method for sparse conditional density estimation in moderately large dimensions}
\maketitle
\textbf{\abstractname:} In this paper, we consider the problem of estimating a conditional density in moderately large dimensions. Much more informative than regression functions, conditional densities are of main interest in recent methods, particularly in the Bayesian framework (studying the posterior distribution, finding its modes...). Considering a recently studied family of kernel estimators, we select a pointwise multivariate bandwidth by revisiting 
the greedy algorithm \Rodeo{} (Regularisation Of Derivative Expectation Operator). The method addresses several issues: being greedy and computationally efficient by an iterative procedure, avoiding the curse of high dimensionality under some suitably defined sparsity conditions by early variable selection during the procedure, converging at a quasi-optimal minimax rate.\\

%

\section{Introduction}
\subsection{Motivations}
In this paper, we consider the problem of the conditional density estimation. 
We observe a $n$-sample of a couple $(X,Y)$
, in which $Y$ is the vector of interest while $X$ gathers auxiliary variables. 
We denote $d$ the joint dimension. In particular we are interested in the inference of the $d$-dimensional conditional density $f$ of $Y$ conditionally to $X$.
\\
There is a growing demand for methods of conditional density estimation in a wide spectrum of applications such as Economy\citep{HRL04}, Cosmology {\citep{IzbickiLee16}}, Medicine \citep{takeuchi09}, Actuaries \citep{efromovich10a}, Meteorology \citep{JeonTaylor} among others. It can be explained by the double role of the conditional density estimation: deriving the underlying distribution of a dataset and determining the impact of the vector $X$ of auxiliary variables on the vector of interest $Y$.
In this aspect, the conditional density estimation is richer than both the unconditional density estimation and the regression problem.
In particular, in the regression framework, only the conditional mean $\E{Y\vert X}$ are estimated instead of the full conditional density, which can be especially poorly informative in case of an asymmetric or multi-modal conditional density. 
Conversely, from the conditional density estimators, one can, \textit{e.g.}, derive the conditional quantiles \citep{TLSS06} or give accurate predictive intervals \citep{FLCLY02}. 
Furthermore, since the posterior distribution in the Bayesian framework is actually a conditional density, the present paper also offers an alternative method to the ABC methodology (for Approximate Bayesian Computation) \citep{beaumont02,MPR12, BCG15} in the case of an intractable-yet-simulable model.\\
The challenging issue in conditional density estimation is to circumvent the "curse of dimensionality".
The problem 
 is twofold: theoretical and practical.
 In theory, it is stigmatized by the minimax approach, stating that in a $d$-dimensional space the best convergence rate for the pointwise risk over a $p$-regular class of functions is $\mathcal{O}(n^{-\frac{p}{2p+d}})$: 
in particular, the larger is $d$, the slower is the rate.
In practice, the larger the dimension is, the larger the sample size is needed to control the estimation error. 
In order to maintain reasonable running times in moderately large dimensions, 
methods have to be designed especially greedy. 
\\
Furthermore, one interesting question is  how to retrieve the eventual \textit{relevant} components in case of sparsity structure on the conditional density
$f$. 
For example, if we have at disposal plenty of auxiliary variables without any indication on their dependency with our vector of interest $Y$, the ideal procedure will take in input the whole dataset and still achieve a running time and a minimax rate as fast as if only the relevant components were given and considered for the estimation. More precisely, two goals are simultaneously addressed : converging at rate $\mathcal{O}(n^{-\frac{2p}{2p+r}})$ with $r$ the relevant dimension, \textit{i.e.} the number of components that influence the conditional density $f$, and  
detect the irrelevant components at an early stage of the procedure in order to afterwards only work on the relevant data and thus speed up the running time.
\subsection{Existing methodologies}
Several nonparametric methods have been proposed to estimate conditional densities: kernel density estimators \citep{rosenblatt69,HBG96,BLR16} and various methodologies for the selection of the associated bandwidth \citep{BH01, FYim04,HRL04}; local polynomial estimators \citep{FYT96, HY02}; projection series estimators \citep{efromovich99, efromovich07}; piecewise constant estimator \citep{GK07, sart17}; copula \citep{faugeras09}. 
But while most of the aforementioned works are only defined for bivariate data or at least when either $X$ or $Y$ is univariate, they are also computationally intractable as soon as $d>3$. \\
%
It is in particular the case for the kernel density methodologies (Hall, Racine ,Li 2004, Bertin et al. 2016): they achieve the optimal minimax rate, and even the detection of the relevant components, thanks to an adequate choice of the bandwidth (for the two aforementioned methods by cross validation and Goldenshluger-Lepski methodology), but the computational cost of these bandwidth selections is prohibitive even for moderate sizes of $n$ and $d$. 
To the best of our knowledge, only two kernel density methods have been proposed to 
handle large datasets.
\citep{HGI10} propose a fast method of approximated cross-validation, based on a dual-tree speed-up, but they do not establish any rate of convergence and only show the consistency of their method.
For scalar $Y$, \citep{FPYZ09} proposed to perform a prior step of dimension reduction on $X$ to bypass the curse of dimensionality, 
then they estimate the bivariate approximated conditional density by kernel estimators. 
But the proved convergence rate $n^{-\frac13}$ is not the optimal minimax rate  $n^{-\frac{3}{8}}$ for the estimation of a bivariate function of assumed regularity $3$. 
Moreover, the step of dimension reduction restricts the dependency of $X$ to a linear combination of its components, which may induce a significant loss of information.
\\
Projection series methods for scalar $Y$ have also been proposed. 
\citep{efromovich10b} extends his previous work \citep{efromovich07} 
to a multivariate $X$. Theoretically the method achieves an oracle inequality, thus the optimal minimax rate. Moreover it performs an automatic dimension reduction on $X$ when there exists a smaller intrinsic dimension.
To our knowledge, it is the only method which addresses datasets of dimension larger than $3$ with reasonable running times and does not pay its numerical performance with non optimal minimax rates. 
However the computation cost is prohibitive when both $n$ and $d$ are large.
More recently, Izbicki and Lee have proposed two methodologies using orthogonal series estimators \citep{IzbickiLee16, IzbickiLee17}.
The first method is particularly fast and can handle very large $X$ (with more than $1000$ covariates).  Moreover the convergence rate adapts to an eventual smaller unknown intrinsic dimension of the support of the conditional density. The second method originally proposes to convert successful high dimensional regression methods into the conditional density estimation, interpreting the coefficients of the orthogonal series estimator as regression functions, which allows to adapt to all kind of figures (mixed data, smaller intrinsic dimension, relevant variables) in function of the regression method. 
However both methods converge slower than the optimal minimax rate. Moreover their optimal tunings depend in fact on the unknown intrinsic dimension. 
\\
For multivariate $X$ and $Y$, \citep{OtneimTjostheim17} propose a new semiparametric method, called Locally Gaussian Density Estimator: they rewrite the conditional density as a product of a function depending on the marginal distribution functions (easily estimated since univariate, then plug-in), and a term which measures the dependency between the components, which is approximated by a centred Gaussian whose covariance is parametrically estimated. Numerically, the methodology seems robust to addition of covariates of $X$ independent of $Y$, but it is not proved. Moreover they only establish the asymptotic normality of their method.  

\subsection{Our strategy and contributions}
The challenge in this paper is to handle large datasets, thus we assume at our disposal a sample of large size $n$ and of moderately large dimension. 
Then our work is motivated by the following three objectives:
\begin{enumerate}[label=(\roman*)]
\item achieving the optimal minimax rate (up to a logarithm term);
\item \label{pointGreedy} being greedy, meaning that the procedure must have reasonable running times for large $n$ and moderately large dimensions, in particular when $d>3$ ;
\item \label{pointSparse} adapting to a potential sparsity structure of $f$. More precisely, in the case where $f$ locally depends only on a number $r$ of its $d$ components, $r$ can be seen as the local \textit{relevant} dimension. Then the desired convergence rate has to adapt to the unknown relevant dimension $r$: under this sparsity assumption, the benchmark for the estimation of a $p$-regular function is to achieve a convergence rate of the order $\mathcal{O}(n^{-\frac{2p}{2p+r}})$, which is the optimal minimax rate if the relevant components were given by an oracle.
\end{enumerate} 
\noindent
Our strategy is based on kernel density estimators.  
The considered family 
has been recently introduced and studied in \citep{BLR16}. This family is especially designed for conditional densities and
is better adapted for the objective \ref{pointSparse} than the intensively studied estimator built as the ratio of a kernel estimator of the joint density over one of the marginal density of $X$. 
 For example, a relevant component for the joint density and the marginal density of $X$ may be irrelevant for the conditional density and it is the case if a component of $X$ is independent of $Y$. Note though that many more cases of irrelevance exist since we define the relevance as a local property.\\
The main issue with kernel density estimators is the selection of the bandwidth $h\in\R_+^d$, and in our case, we also want to complete the objective \ref{pointGreedy}, since the pre-existing methodologies of bandwidth selection does not satisfy this restriction and thus cannot handle large datasets. 
In this paper, it is performed by an algorithm we call \CDRodeo, which is derived from the algorithm \Rodeo\citep{LW08,LLW07}, which has respectively been applied for the regression and the unconditional density estimation. The greediness of the algorithm allows us to address datasets of large sizes while keeping a reasonable running time (see Section \ref{sectionComplexity} for further details). We give a simulated example with a sample of size $n=10^5$ and of dimension $d=5$ in Section \ref{sectSimu}. Moreover, \Rodeo-type algorithms ensure an early detection of irrelevant component, and thus achieve the objective \ref{pointSparse} while improving the objective \ref{pointGreedy}.\\
From the theoretical point of view, if the regularity of $f$ is known, our method achieves an optimal minimax rate (up to a logarithmic factor), which is adaptive to the unknown sparsity of $f$. The last property is mostly due to the \Rodeo-type procedures. The improvement of our method in comparison to the paper \citep{LLW07} which estimates the \textit{unconditional} density with \Rodeo is twofold.
First, our result is extended to any regularity $p\in\mathbb{N}_{>0}$, whereas \citep{LLW07} fixed $p=2$.
Secondly, our notion of relevance is both less restrictive and more natural. In \citep{LLW07}, they studied the $L_2$-risk of their estimator, therefore they have to consider a notion of global relevance, whereas we consider a pointwise approach, which allows us to define a local property of relevance, which can be applied to a broader class of functions. 
Moreover, their notion of relevance is not intrinsic to the unknown density, but in fact depends on a tuning of the method, a prior chosen \textit{baseline density} which has no connexion with the density,  
which limits the interpretation of the \textit{relevance}. 
%
%
%
%
%
\subsection{Overview}
Our paper is organized as follows.
We introduce the \CDRodeo method in \Cref{sectionMethod}. 
The theoretical results are in \Cref{sectionThm}, in which we specify the assumptions and the tunings of the procedure from which are derived the convergence rate and the complexity cost of the method.
A numerical example is presented in \Cref{sectSimu}. 
The proofs are in the last section.
\section{\CDRodeo method}\label{sectionMethod}
Let $W_1,\dots,W_n$ be a sample of a couple $(X,Y)$ of multivariate random vectors: 
for $i=1,\dots,n$, $$W_i=(X_i,Y_i),$$
with $X_i$ valued in $\R^{d_1}$ and $Y_i$ in $\R^{d_2}$. We denote $d:=d_1+d_2$ the joint dimension.\\
 We assume that the marginal distribution of $X$ and the conditional distribution of $Y$ given $X$ are absolutely continuous with respect to the Lebesgue measure, and we define $f:\R^d\rightarrow\R$ such as for any $x\in\R^{d_1}$, $f(x,\cdot)$ is the conditional density of $Y$ conditionally to $X=x$. We denote $f_X$ the marginal density of $X$.\\
 Our method estimates $f$ pointwisely : let us fix $w=(x,y)\in\R^d$ the point of interest.

\paragraph{Kernel estimators.}
Our method is based on kernel density estimators. 
More specifically, we consider the family proposed in \citep{BLR16}, which is especially designed for the conditional density estimation. 
Let $K:\R\rightarrow\R$ be a kernel function, \textit{ie:} $\int_{\R} K(t)dt =1$, then for any bandwidth $h\in(\mathbb{R}^*_+)^d$, the estimator of $f(w)$ is defined by:
\begin{equation}\label{eDef fh}
\f{h}:=\frac{1}{n} \sum\limits_{i=1}^n \frac{1}{\fX\left(X_i\right)} \prod\limits_{j=1}^{d}h_j^{-1}K\left(\frac{{w}_j-W_{ij}}{h_j}\right),
\end{equation}
where $\fX$ is an estimator of $f_X$, built from another sample $\X$ of $X$. 
We denote by $\nX$ the sample size of $\X$. The choices of $K$ and $\fX$ are specified later (see section \ref{sectionEstimatorChoice}). 
\paragraph{Bandwidth selection.}
In kernel density estimation, selecting the bandwidth is a critical choice which can be viewed as a bias-variance trade-off. 
 In \citep{BLR16}, it is performed by the Goldenshluger-Lepski methodology (see \citep{GL11}) and requires an optimization over an exhaustive grid of couples $(h,h')$ of bandwidths, which leads to intractable running time when the dimension exceeds $3$ (and large dataset).\\
 That is why we focus in a method which excludes optimization over an exhaustive grid of bandwidths to rather propose a greedy algorithm 
  derived from the algorithm \Rodeo. 
 First introduced in the regression framework \citep{LW06,LW08}, a variation of \Rodeo was proposed in \citep{LLW07} for the 
 density estimation. Our method  we called \CDRodeo{} (for Conditional Density \Rodeo) addresses the more general problem of conditional density estimation.\\
%
Like \Rodeo (which means Regularisation Of Derivative Expectation Operator), the \CDRodeo algorithm  generates an iterative path of decreasing bandwidths, based on tests on the partial derivatives of the estimator with respect to the components of the bandwidth.
Note that the greediness of the procedure leans on the selection of this path of bandwidths, 
which enables us to address high dimensional problems of functional inference. 
\\
\noindent Let us be more precise: we take a kernel $K$ of class $\mathcal{C}^1$ and consider the statistics $Z_{hj}$ for $h\in(\R_+^*)^d$ and $j=1:d$, defined by:
$$Z_{hj}:= \frac{\partial}{\partial h_j} \hat{f}_h(w).$$
 $Z_{hj}$ is easily computable, since it can be expressed by:
\begin{equation}
Z_{hj}=\frac{-1}{n h_j^2}
 \sum\limits_{i=1}^n \frac{1}{\fX(X_i)} J\left(\tfrac{w_j-W_{ij}}{h_j}\right)
\prod\limits_{k\neq j}^{d}h_k^{-1}K\left(\frac{{w}_k-W_{ik}}{h_k}\right),
\end{equation}
where $J:\R\rightarrow\R$ is the function defined by:
\begin{equation}\label{eJdef}
 t\mapsto K(t) + tK'(t).
\end{equation}

\begin{algorithm*}
\caption{\CDRodeo algorithm \label{algo}}
\begin{enumerate}
\item \textit{Input:} the point of interest $w$, the data $W$, $\beta\in (0,1)$ the bandwidth decreasing factor, $h_0>0$ the bandwidth initialization value, a parameter $a>1$.
\item \textit{Initialization:}{\begin{enumerate}		
		\item Initialize the bandwidth: \textit{for }$j=1:d$, $h_j\leftarrow h_0$.
		\item Activate all the variables: $\mathcal{A}\leftarrow\lbrace 1,\dots,d \rbrace$.
	\end{enumerate}
	}
\item {\textit{While} ($\mathcal{A}\neq\emptyset $) \& ($\prod\limits_{k=1}^d h_k\geq \frac{\log n}{n}$): 
\begin{itemize}
	\item[] {\textit{for all $j\in\mathcal{A}$:}\begin{enumerate}
		\item Update $Z_{hj}$ and $\seuil{hj}$.
		\item {\textit{If} $\vert Z_{hj}\vert \geq \seuil{hj}$: update $h_j\leftarrow\beta h_j$. \\
		\textit{else:} remove $j$ from $\mathcal{A}$.}
	\end{enumerate}
	}
\end{itemize}
}
\item \textit{Output:} $h$ (and $\hat{f}_h(w)$).
\end{enumerate}
\end{algorithm*}
\noindent The details of the \CDRodeo procedure are described in \textbf{Algorithm \ref{algo}} and can be summed up in one sentence: for a well-chosen threshold $\seuil{hj}$ (specified in \Cref{sectionRodeoParameter}),
the algorithm performs at each iteration the test $|Z_{hj}|>\seuil{hj}$  to determine if the component $j$ of the current bandwidth must be shrunk or not. It can be interpreted by the following principle: 
the bandwidth of a kernel estimator quantifies within which distance of the point of interest $w$ and at which degree an observation $W_i$ helps in the estimation. Heuristically, the larger the variation of $f$ is, the smaller the bandwidth is required for an accurate estimation. The statistics $Z_{hj}=\frac{\partial}{\partial h_j} \f{h}$ are used as a proxy of $\frac{\partial}{\partial w_j}f(w)$ to quantify the variation of $f$ in the direction $w_j$.
 Note in particular that since the partial derivatives vanish for irrelevant components, this bandwidth selection leads to an implicit variable selection, and thus to avoid the curse of dimensionality under sparsity assumptions.


\section{Theoretical results}\label{sectionThm}
This section gathers the theoretical results of our method.
\subsection{Assumptions}\label{sectionAssp}
We consider $K$ a compactly supported kernel. For any bandwidth $h\in(\R_+^*)^d$, we define the neighbourhood $\mathcal{U}_h(u)$ of $u\in\R^{d'}$  (typically, $u=x$ or $w$, and $d'=d_1$ or $d$) as follows: 
 $$\mathcal{U}_h(u):=\left\lbrace u'\in\mathbb{R}^{d'}: \forall j=1:d', u'_j=u_j-h_jz_j, \text{with } z\in\left(\supp(K)\right)^{d'}\right\rbrace.$$
Then we denote the \CDRodeo{} initial bandwidth $h^{(0)}=\left( \tfrac{1}{\log n},\dots,\tfrac{1}{\log n}\right)$ and for short, $\mathcal{U}_n(u):= \mathcal{U}_{h^{(0)}}(u)$.\\
  We also introduce the notation $\Vert\cdot\Vert_{\infty,\ \mathcal{U}}$ for the supremum norm over a set $\ \mathcal{U}$.\\
  ~\\
  The following first assumption ensures a certain amount of observations in the neighbourhood of our point of interest $w$.   

 \begin{assp}[$f_X$ bounded away of $0$]\label{AfXmin}
We assume $\delta := \inf\limits_{u\in \Un} f_X(u)>0$.
\end{assp} 
\noindent  Note that if the neighbourhood $\Un$ does not contain any observation $X_i$, the estimation of the conditional distribution of $Y$ given the event $X=x$ is obviously intractable. \\

\noindent The second assumption specifies the notions of "sparse function" and "relevant component", under which the curse of high dimensionality can be avoided.
\begin{assp}[Sparsity condition]\label{Afsparse}
There exists a subset $\mathcal{R}\in\lbrace1,\dots,d\rbrace$ such that for any fixed $\lbrace z_j\rbrace_{j\in\mathcal{R}}$, the function $\lbrace z_k\rbrace_{k\in\mathcal{R}^c}\mapsto f(z_1,\dots,z_d)$ is constant on $\UUn$.
\end{assp}
\noindent In other words, if we denote  $r$ the cardinal of $\mathcal{R}$, \Cref{Afsparse} means that $f$ locally depends on only $r$ of its $d$ variables. We call \textit{relevant} any component in $\mathcal{R}$.
The notion of relevant component depends on the point where $f$ is estimated. For example, a component $w_j$ which behaves as $\ind_{[0,1]}(w_j)$ in the conditional density is only relevant in the neighbourhood of $0$ and $1$.
Note that this local property addresses a broader class of functions, which extends the application field of \Cref{thmRodeo} and improves the convergence rate of the method.\\
~\\
Finally, the conditional density is required to be regular enough. 
 \begin{assp}[Regularity of $f$]  \label{Afreg}
There exists a known integer $p$ such that $f$ is of class $\mathcal{C}^p$ on $\UUn$ and such that $\partial_j^p f(w)\neq 0$ for all $j\in\rel$.
\end{assp}
\noindent 


\subsection{Conditions on the estimator of $f_X$}\label{sectionEstimatorChoice}
Given the definition of the estimator \eqref{eDef fh}, we need an estimator $\fX$ of $f_X$. 
\paragraph{If $f_X$ is known.} 
We take $\fX \equiv f_X$.
This case is not completely obvious. In particular, it tackles the case of \textit{un}conditional
 density estimation, if we set by convention $d_1=0$ and $f_X\equiv 1$. 
 
 \paragraph{If $f_X$ is unknown.} 
We need an estimator $\fX$ which satisfies the following two conditions:
  \begin{enumerate}[label=(\roman*)]
  \item a positive lower bound: $\dX:=\inf\limits_{u\in \Un} \fX(u)>0$ \label{fXtildemin}
  \item a concentration inequality in local sup norm: \label{fXtildeAccuracy} there exists a constant $M_{X}>0$ such that:  $$\PP{\sup\limits_{u\in\Un}\left\vert \frac{f_X(u)-\fX(u)}{\fX(u)}\right\vert > M_{X} \frac{(\log n)^{\frac{d}2}}{n^{\frac12}}}\leq  \exp(-(\log n)^{\frac54}).$$
  \end{enumerate}
  The following proposition proves these conditions are feasible. Furthermore, the provided estimator of $f_X$ (see the proof in \Cref{sectProoffXtilde}) is easily implementable and does not need any optimisation.
\begin{prop}\label{propfXtilde} Given a sample $\X$ with same distribution as $X$ and of size $\nX=n^c$ with  $c>1$,
if $f_X$ is of class $\mathcal{C}^{p'}$ with $p'\geq \frac{d_1}{2(c-1)}$, there exists an estimator $\fX$ which satisfies \ref{fXtildemin} and \ref{fXtildeAccuracy}.
\end{prop}
%
%

\subsection{\CDRodeo{} parameters choice.} \label{sectionRodeoParameter}
\paragraph{Kernel $K$.}\label{KC} 
 We choose the kernel function $K:\R\rightarrow \R$ of class $\mathcal{C}^1$, with compact support and of order $p$, \textit{i.e.}: for $\ell=1,\dots,p-1$, $\int_{\R} t^\ell K(t)dt=0$, and $\int_{\R} t^p K(t)dt\neq 0$.\\
%
Note that considering a compactly supported kernel is fundamental for the local approach. In particular, it relaxes the assumptions by restricting them to a neighbourhood of $w$.\\
Taking a kernel of order $p$ is usual for the control of the bias of the estimator.
\paragraph{Parameter $\beta$.} Let  $\beta\in(0,1)$ be the decreasing factor of the bandwidth. The larger $\beta$, the more accurate the procedure, but the longer the computational time. From the theoretical point of view, it remains of little importance, as it only affects the constant terms. 
In practice, we set it close to $1$.

\paragraph{Bandwidth initialization.} We recall that we set $h_0:=\frac{1}{\log n}$ (and the initial bandwidth as $\left( \tfrac{1}{\log n},\dots,\tfrac{1}{\log n}\right)$). 

\paragraph{Threshold $\seuil{h,j}$.} For any bandwidth $h\in(\R_+^*)^d$ and for $j=1:d$, we set the threshold as follows:
\begin{equation}
\seuil{hj}:= \Cl \sqrt{\frac{(\log n)^{a}}{n h_j^2 \prod_{k=1}^d h_k}},
\end{equation}
with
$\Cl:= 4\Vert J\Vert_2\Vert K\Vert_2^{d-1}$ (where $J$ is defined in \eqref{eJdef}) and $a>1$. The expression is obtained by using concentration inequalities on $Z_{hj}$. For the proof, the parameter $a$ has to be tuned such that:\begin{equation}\label{eConda}
(\log n)^{a-1}>\frac{\Vert f\Vert_{\infty, \UUn}}{\delta},
\end{equation}
which is satisfied for $n$ large enough. The influence of this parameter is discussed in the next section, once the theoretical results are stated.\\
~\\
Hereafter, unless otherwise specified, the parameters are chosen as described in this section.

\subsection{Mains results}\label{sectionResults}
Let us denote $\hat{h}$ the bandwidth selected by \CDRodeo.
In  \Cref{thmRodeo}, we introduce a set $\HGP$ of bandwidths which contains $\hat{h}$ with high probability, which leads to an upper bound of the pointwise estimation error with high probability. 
In \Cref{cor}, we deduce the convergence rate of \CDRodeo from \Cref{thmRodeo}. \\
More precisely, in \Cref{thmRodeo}, we determine lower and upper bounds (with high probability) for the stopping iteration of each  bandwidth component. 
 We set:
\begin{equation}
\preA
:=\frac{1}{(2p+r)\log\frac1\beta} \log\left(\frac{n}{\Ctau(\log n)^{2p+d+a}}\right),
\label{eDefpreA}
\end{equation}
and 
\begin{equation}
\postA:= \preA + \frac{\log \left(\CT^{-1}\right)}{(2p+1)\log\frac1\beta},
\end{equation}
 where 
$$\Ctau:=\left(\frac{4(p-1)!\Cl}{ \left(\min\limits_{j\in\rel}\partial_j^pf(w)\right)\int_{\R}t^p K(t)dt  }\right)^2, \,
\CT:=\left(\frac{\min\limits_{j\in\rel}\vert\partial_j^pf(w)\vert}{24\max\limits_{j\in\rel}\vert\partial_j^pf(w)\vert}\right)^2.$$\\
Then we define the set of bandwidths $\HGP$ by: $$\HGP:=\left\lbrace h\in\R_+^d : h_j=\frac{\beta^{\theta_j}}{\log n},\text{ with }\theta_j\in\lbrace\lfloor\preA\rfloor+1, \dots, \lfloor\postA\rfloor\rbrace \text{ if }j\in\rel, \text{ else  }\theta_j=0\right\rbrace .$$

\begin{theo} \label{thmRodeo}  Assume that $\fX$ satisfies Conditions \ref{fXtildemin} and \ref{fXtildeAccuracy} of section \ref{sectionEstimatorChoice} and  \Cref{AfXmin,Afreg,Afsparse} are satisfied. 
Then, the bandwidth $\hat{h}$ selected by \CDRodeo  belongs to $\HGP$  with high probability. 
More precisely, for any $q>0$ and for $n$ large enough: 
\begin{equation}\label{eCDRBandwidthSelection}
\mathbb{P}\left( \hat{h}\in\HGP \right)\geq 1- n^{-q}.
\end{equation}
Moreover, with probability larger than $1- 2n^{-q}$, the \CDRodeo estimator $\hat{f}_{\hat{h}}(w)$  verifies:
\begin{equation}\label{eCDRerror}
\left\vert \f{\hat{h}}-f(w)\right\vert 
\leq \text{C} (\log n)^{\frac{p}{2p+r}(d-r+a)} n^{-\frac{p}{2p+r}} 
\end{equation}
with $$\text{C}:= 2r \Ctau^{\frac{p}{2p+r}}\int_{t\in\R} \vert  \frac{t^p}{p!} K(t) \vert dt\times \max\limits_{k\in\rel} \Vert \partial_k^p f\Vert_{\infty,\UUn} + 4\Vert K\Vert^d_2 \Vert f\Vert_{\infty,\UUn}^{\frac12}\delta^{-\frac12}\CT^{\frac{-r}{2(2p+1)}}\Ctau^{\frac{-r}{2(2p+r)}}.$$
\end{theo}

\begin{cor}\label{cor}
Under the assumptions of \Cref{thmRodeo}, for any $q\geq1$:
\begin{equation*}
\left(\E{\left\vert\f{\hat{h}}-f(w)\right\vert^q}\right)^{1/q}\leq \text{C}(\log n)^{\frac{p}{2p+r}(d-r+a)} n^{-\frac{p}{2p+r}}
+o\left(n^{-1}\right).
\end{equation*}
\end{cor}
\noindent \Cref{cor} presents a generalization of the previous works on \Rodeo 
\citep{LW08} and \citep{LLW07} whose results are restricted to the regularity $p=2$ and to simpler problems, namely regression and density estimation.\\
We compare the convergence rate of \CDRodeo with the optimal minimax rate. 
In particular, our benchmark is the pointwise minimax rate, which is of order $\mathcal{O}\left(n^{-\frac{p}{2p+d}}\right)$,  for the problem of $p$-regular $d$-dimensional density estimation, obtained by \citep{DL92}.\\
\noindent  
%
Without sparsity structure ($r=d$), \CDRodeo achieves the optimal minimax rate, up to a logarithmic factor.
The exponent of this factor depends on the parameter $a$. For the proofs, we need $a>1$ in order to satisfy \eqref{eConda}, but if an upper bound (or a pre-estimator) of $\frac{\Vert f\Vert_{\infty, \UUn}}{\delta}$ were known, we could obtain the similar result with $a=1$ and a modified constant term.  
Note that the logarithmic factor is a small price to pay for a computationally-tractable procedure for high-dimensional functional inference, in particular see section \ref{sectionComplexity} for the computational gain of our procedure. 
 
\noindent
%
Under sparsity assumptions, we avoid the curse of high dimensionality and our procedure achieves 
the desired rate $n^{-\frac{p}{2p+r}}$ (up to a logarithmic term), which is optimal if the relevant components were known. 
Note that some additional logarithmic factors could be unavoidable due to 
the unknown sparsity structure, which needs to be estimated. Identifying the exact order of the logarithm term in the optimal minimax rate for the sparse case remains an open challenging question.

\subsection{Complexity}\label{sectionComplexity}
We now discuss the complexity of \CDRodeo{}
without taking into account the pre-computation cost of $\fX$ at the points $X_i$, $i=1:n$ (used for computing the $Z_{hj}$), but a fast procedure for $\fX$ is required, to avoid losing \CDRodeo computational advantages.\\
For \CDRodeo, the main cost lies in the computation of the $Z_{hj}$'s along the path of bandwidths.
\\
The condition $\prod\limits_{k=1}^d h_k\geq \frac{\log n}{n}$ restricts to at most $\log_{\beta^{-1}} n$  updates of the bandwidth across all components, 
leading to a worst-case complexity of order $\mathcal{O}(d.n\log n)$. \\
But as shown in \Cref{thmRodeo}, with high probability, $\hat{h}\in\HGP$, in which only the relevant components are active after the first iteration. In first iteration, the $Z_{h^{(0)}j}$'s computation costs $\mathcal{O}(d.n)$ operations, while the product kernel enables us to compute the $Z_{hj}$'s in following iteration with only $\mathcal{O}(r.n)$ operations, which leads to the complexity $\mathcal{O}(d.n+r.n\log n)$.\\
In order to grasp the advantage of \CDRodeo{} greediness, we compare its complexity with optimization over an exhaustive bandwidth grid with $\log n$ values for each component of the bandwidth (which is often the case in others methods: Cross validation, Lepski methods...): for each bandwidth of  $(\log n)^d$-sized grid, the computation of a  statistic from the $d.n$-sized dataset needs at least $\mathcal{O}(d.n)$ operation, which leads to a complexity of order $\mathcal{O}(d.n(\log_{\beta^{-1}}  n)^d)$. Using the parameters used in the simulated example in section \ref{sectSimu} 
($n=2.10^5$, $d=5$, $r=3$, $\beta=0.95$), the ratio of complexities is $\dfrac{d. n(\log n)^d}{r.n\log n}\approx 5.10^9$, 
 and even without sparsity structure: $\dfrac{d. n(\log n)^d}{d.n\log n}\approx 3.10^9$. It means that  \CDRodeo{} run is a billion times faster on this data set.
 

\section{Simulations}\label{sectSimu}
\begin{figure}[H]
\includegraphics[width=\linewidth]{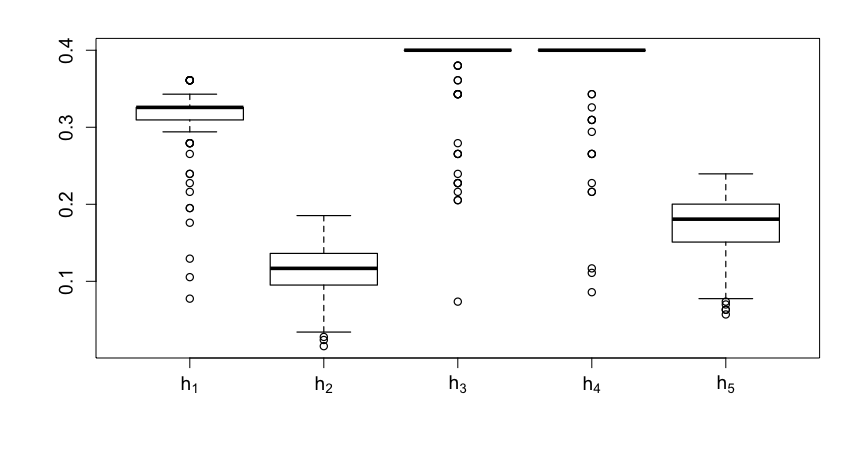}
\vspace{-5em}
\caption{ Boxplots of each component of $200$ \CDRodeo selected bandwidths at the point $w=(0,1,0,0,1)$.}\label{figBoxpBandwidth}
\vspace{-0.5em}
\end{figure}

\begin{figure}[h]
\includegraphics[width=0.5\linewidth]{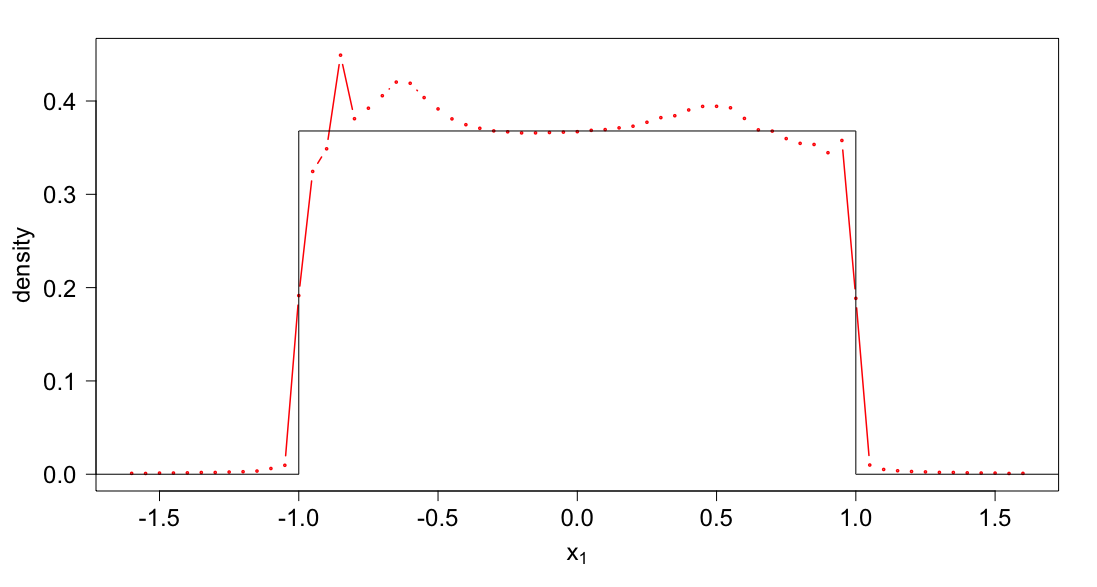}
\includegraphics[width=0.5\linewidth]{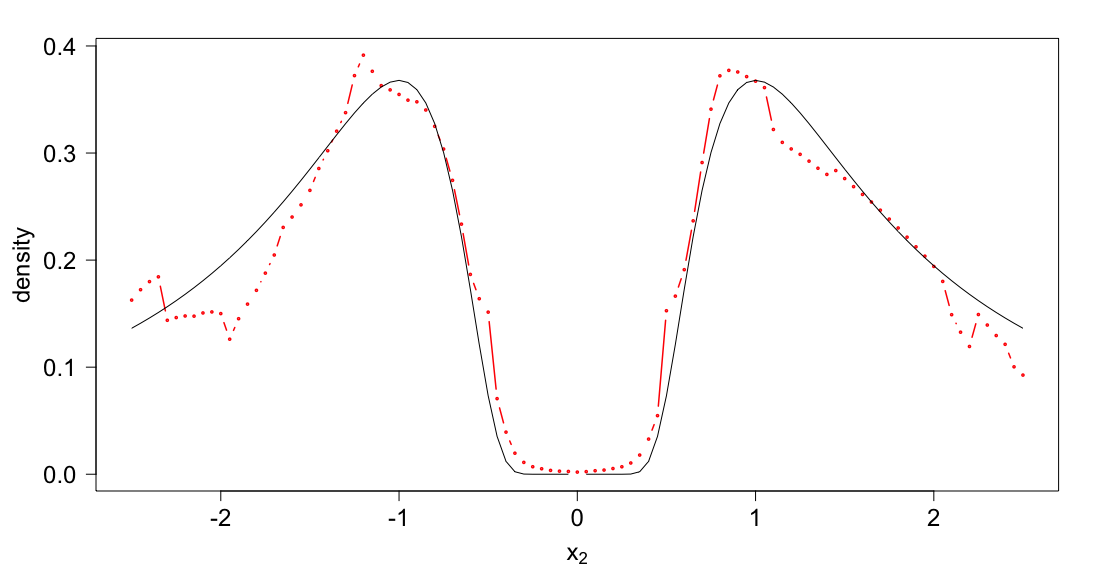}
\includegraphics[width=0.5\linewidth]{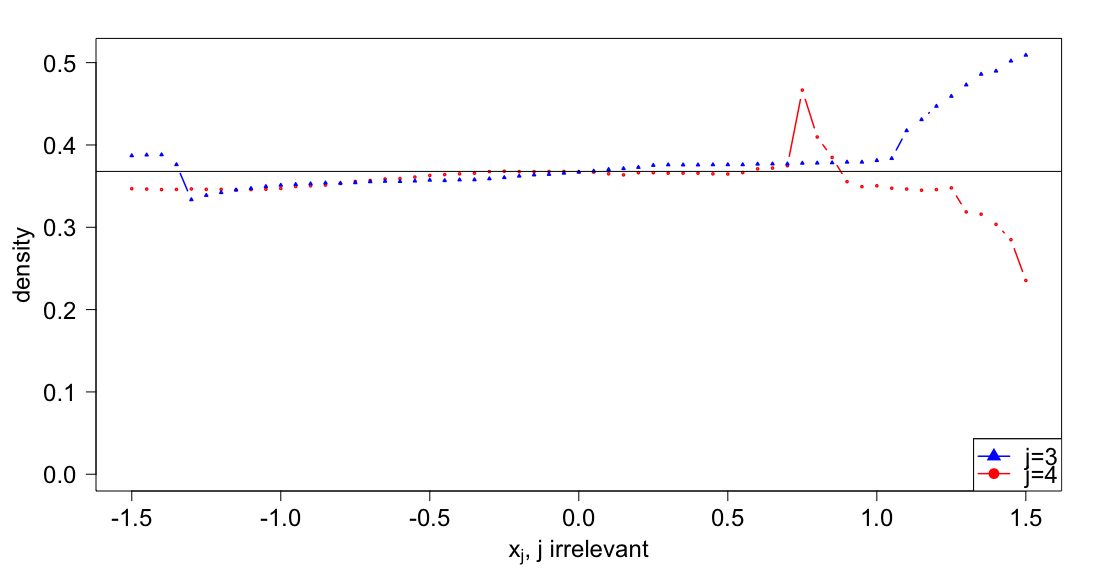}
\includegraphics[width=0.5\linewidth]{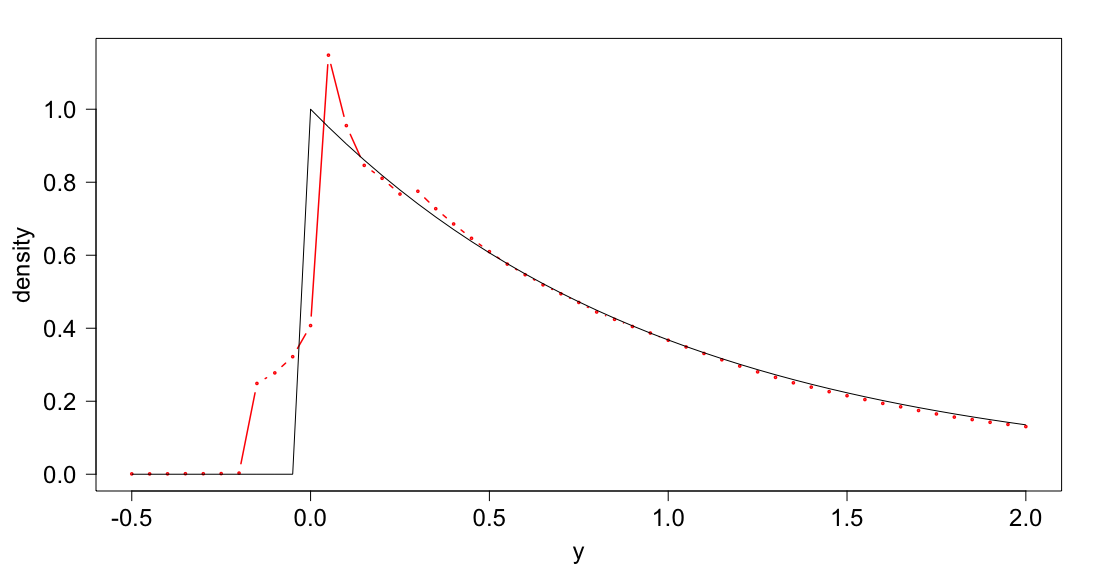}
\vspace{-2.5em}
\caption{\CDRodeo{} estimator (red or blue dashed lines) VS the true density (black solid line) in function of each component, the others component being fixed following $(x,y) = (0,1,0,0,1)$.}
\label{figDimParDim}
\vspace{-0.5em}
\end{figure}

In this section, we test the practical performances of our method. In particular, we study \CDRodeo{} on a 5-dimensional example.
The major purpose of this section is to assess if the numerical performances of our procedure. 
Let us describe the example. We set $d_1=4$ and $d_2=1$ and simulate an i.i.d sample $\lbrace(X_i,Y_i)\rbrace_{i=1}^n$ with the following distribution: for any $i=1,\dots,n$:\begin{itemize}
\item[-] the first component $X_{i1}$ of $X_i$  follows a uniform distribution on $[-1,1]$,
\item[-] the other components $X_{ij}$, $j=2:4$, are independent standard normal and are independent of $X_{i1}$,
\item[-] $Y_i$ is independent of $X_{i1}$, $X_{i3}$ and  $X_{i4}$ and the conditional distribution of $Y_i$ given $X_{i2}$ is exponential with survival parameter $X_{i2}^{2}$.
\end{itemize}   
The estimated conditional density function is then defined by: 
$$f:
(x,y)\mapsto \ind_{[-1,1]}(x_1) \frac{1}{x_2^2}e^{-\frac{y}{x_2^2}}.$$ 
This example enables us to test several criteria: sparsity detection, behaviour 
when fonctions are not continuous, bimodality estimation, robustness when $f_X$ takes small values. \\
In the following simulations, if not stated explicitly otherwise, \Rodeo{} is run with 
sample size $n=200,000$, product Gaussian kernel, initial bandwidth value $h_0=0.4$, bandwidth decreasing factor $\beta=0.95$ and parameter $a=1.1$ and $\fX\equiv f_X$.\\
Figure \ref{figBoxpBandwidth} illustrates \CDRodeo{} bandwidth selection. In which, the boxplots of each selected bandwidth component are built from 200 runs of \CDRodeo{} at the point $w=(0,1,0,0,1)$. 
This figure reflects the specificity of \CDRodeo to capture the relevance degree of each component, and 
one could compare it with variable selection (as done in \citep{LW08}).
 The components $x_3$ and $x_4$ are irrelevant and for this point of interest, the components $x_2$ and $y$ are clearly relevant while the component $x_1$  is barely relevant as $f$ is constant in the direction $x_1$ in near neighbourhood of $x_1=0$. 
 As expected, the irrelevant $h_3$ and $h_4$ are mostly deactivated at the first iteration, while the relevant $h_2$ and $h_5$ are systematically shrunk. The relevance degree of $x_1$ is also well detected as the values of $h_1$ are smaller than $h_0$, but significantly larger than $h_2$ and $h_5$.\\
~\\%
Figure \ref{figDimParDim} gives \CDRodeo{} estimation of $f$ from one $n$-sample. 
The function $f$ is well estimated.
In particular, irrevance, jumps and bi-modality are features which are well detected by our method.
As expected, main estimation errors are made on points of discontinuity for $x_1$ and $y$ or at the boundaries for $x_2$,  $x_3$ and  $x_4$. 
Note that the $f_X$ values are particularly small at the boundaries of the plots in function of $x$, leading to lack of observations for the estimation.
Note however that null value for $f_X$ does not deteriorate the estimation (cf top left plot), 
since the estimate of $f$ vanishes automatically when there is no observation near the point of interest.

\paragraph{Running time.} The simulations are implemented in R on a Macintosh laptop with a $3,1$ GHz Intel Core i7 processor. In the Figure \Cref{figBoxpBandwidth}, the $200$ runs of \CDRodeo{} take $2952.735$ seconds (around $50$ minutes), or $14.8$ seconds per run.

\section{Proofs}
 We first give the outlines of the proofs in \Cref{ssectOutlineProofs}. To facilitate the lecture of the proof,  we have divided the proofs of the main results (\Cref{propfXtilde}, \Cref{thmRodeo} and \Cref{cor}) into intermediate results which are stated in \Cref{ssect IntermediateLemma} and proved in \Cref{ssectProofIntermediate}. The proof of the main results are in \Cref{ssectProofMainResults}. 
 
\subsection{Outlines of the proofs}\label{ssectOutlineProofs}
We first prove \Cref{propfXtilde} by constructing an estimator of $f_X$ with the wanted properties. In this proof, we use some usual properties of a kernel density estimator (control of the bias, concentration inequality), which are gathered in \Cref{lmfXK}.\\
\Cref{thmRodeo} states two results: the bandwidth selection \eqref{eCDRBandwidthSelection} and the estimation error of the procedure \eqref{eCDRerror}.
For the proof of the bandwidth selection \eqref{eCDRBandwidthSelection}, \Cref{propRod} mades explicit the highly probable behaviour of \CDRodeo along a run,  and thus the final selected bandwidth. In particular, the proof leans on an analysis of $Z_{hj}$, which is made in two steps. We first consider $\Zbar{hj}$, a simpler version of $Z_{hj}$ in which we substitute the estimator of $f_X$ by $f_X$ itself, and we detail its behaviour in \Cref{lmZbar}. Then  we control the difference $Z_{hj}- \Zbar{hj}$ (see in 1. of \Cref{lmD}) to ensure $\Zbar{hj}$ behaves like $Z_{hj}$.\\ 
To control the estimation error of the procedure \eqref{eCDRerror}, we similarly analyse $\f{h}$ in two parts: in \Cref{lmfbar}, we describe the behaviour of $\fbar{h}$, the simpler version of $\f{h}$ in which we substitute the estimator of $f_X$ by $f_X$ itself, and in 2. of \Cref{lmD}, we bound the difference $f_{h}- \fbar{h}$. Then the bandwidth selection \eqref{eCDRBandwidthSelection} leads to the upper bound with high probability of the estimation error of $\f{h}$  \eqref{eCDRerror}.\\
Finally, we obtain the expected error of $\f{h}$ stated in \Cref{cor} by controlling the error on the residual event.


\subsection{Intermediate results} \label{ssect IntermediateLemma}
For any bandwidth $\hX\in\R_+^*$, we define 
   the kernel density estimator $\fXK$ by: for any $u\in\R^{d_1}$,
   \begin{equation}\label{eDef fXK}
  \fXK (u):= \frac{1}{\nX.\hX^{d_1}}\sum\limits_{i=1}^{\nX} \prod\limits_{j=1}^{d_1} \KX \left(\frac{u_j-\X_{ij}}{\hX}\right),
  \end{equation}
where $\KX:\R\rightarrow\R$ a kernel which is compactly supported, of class $\mathcal{C}^1$, 
of order $p_X\geq \frac{d_1}{2(c-1)},$ where we recall that $c>1$ is defined by $\nX=n^c$.\\
  We also introduce the neighbourhood  \begin{equation}\label{edefUp}
   \Up:=\lbrace{u'=u-h_Xz : u\in\Un, z\in\supp(\KX)}\rbrace.
  \end{equation}
\begin{lem}[$\fXK$ behaviour]\label{lmfXK}
  We assume $f_X$ is $\mathcal{C}^{p'}$ on $\Up$ with $p'\leq p_X$, then for any bandwidth $\hX\in\R_+^*$,
\begin{enumerate}
\item if we denote $\CbiasX:=\frac{\Vert\KX\Vert_1^{d_1-1}\Vert\cdot^{p'}\KX(\cdot)\Vert_1}{p'!} d_1\max\limits_{k=1:d_1}\Vert \partial_k^{p'}f_X\Vert_{\infty,\Up}$, then
\begin{align*}
\left\Vert \E{\fXK}- f_X\right\Vert_{\infty,\Un}
&\leq \CbiasX h_X^{p'}.
\end{align*}
\item  If the condition 
$$\text{Cond}_X(\hX): \quad \hX^{d_1}
\geq \frac{4\Vert \KX\Vert_{\infty}^{2d_1}}{9\Vert \KX\Vert_2^{2d_1}\Vert f_X\Vert_{\infty,\Up}} \frac{(\log n)^{\frac32}}{\nX}$$ is satisfied, then for $\lambda_X:=\sqrt{\tfrac{4\Vert \KX\Vert_2^{2d_1}\Vert f_X\Vert_{\infty,\Up}}{\hX^{d_1}\nX}(\log n)^{\frac32}}$ and for any $u\in\Un$:
\begin{align*}
\PP{\left\vert\fXK(u)-\E{\fXK(u)}\right\vert>\lambda_X}
\leq 2 \exp\left(- (\log n)^{\frac32}\right).
\end{align*}
\end{enumerate}
\end{lem}

\begin{lem}[$\fbar{h}$ behaviour]\label{lmfbar}
 For any bandwidth $h\in(0,h_0]^d$, and any $i=1:n$, let us denote $\fbar{hi}:= \frac{K_h(w-W_i)}{f_X(X_i)}$. Then,  if $K$ is chosen as in section \ref{KC}, under \Cref{Afreg,Afsparse,AfXmin},
\begin{enumerate}
\item Let $\CEbar:=\Vert f\Vert_{\infty,\UUn}\Vert K\Vert_1^d$. Then
$$\left\vert \E{\fbar{h1}} \right\vert \leq \E{\left\vert\fbar{h1} \right\vert}
\leq \CEbar.$$
Besides, if we denote $\overline{B}_h:=\E{\fbar{h}}-f(w)$ the bias of 
 $\fbar{h}:=\frac1n \sum\limits_{i=1}^n \fbar{hi}$, then:
$$ \vert \overline{B}_h\vert
\leq \Cbiasbar \sum\limits_{k\in\mathcal{R}} h_k^p,$$
with $\Cbiasbar:=  \frac{2\vert \int_{t\in\R} t^p K(t) dt\vert}{p!}\max\limits_{k\in\rel} \vert \partial_k^p f(w)\vert $.
\item Let $\B{h}:=\lbrace \vert \fbar{h}-\mathbb{E}[\fbar{h}]\vert \leq \sigma_h\rbrace$,
 where $\sigma_h:=\Cs\sqrt{ \frac{(\log n)^a}{ n \prod\limits_{k=1}^d h_k}}$ with
$ \Cs=\frac{2\Vert K\Vert^d_2 \Vert f\Vert_{\infty,\UUn}^{\frac12}}{\delta^{\frac12}}$.
If  Cond($h$): $\prod\limits_{k=1}^d h_k\geq \frac{4\Vert K \Vert^{2d}_{\infty}}{9\delta^2\Cs^2} \frac{(\log n)^a}{n}$ is satisfied, then:
 $$\mathbb{P}\left( \B{h}^c \right) \leq 2 e^{-(\log n)^a}$$
 \item Let  $\BA{h}:=\lbrace \vert \frac1n\sum\limits_{i=1}^n \vert\fbar{hi}\vert-\mathbb{E}[\vert\fbar{h}\vert]\vert \leq \CEbar \rbrace$.
 Then 
 $$\mathbb{P}\left( \BA{h}^c \right) \leq 2 e^{-\CgfA n\prod_{k=1}^d h_k},$$
 \end{enumerate}
 with $\CgfA:= \min\left( \frac{\CEbar^2}{\Cs^2} ; \frac{3\delta \CEbar}{4\Vert K \Vert^d_{\infty}} \right)$.
\end{lem}

\begin{lem}[$\bar{Z}_{hj}$ behaviour]\label{lmZbar}
For any $j\in \lbrace 1,\dots,d\rbrace$ and any bandwidth $h\in(0,h_0]^{d}$, we define
\linebreak $\bar{Z}_{hij}:= \frac{1}{f_X\left(X_i\right)} \dfrac{\partial}{\partial h_j}\left(\prod\limits_{k=1}^{d}h_k^{-1}K\left(\frac{{w}_k-W_{ik}}{h_k}\right)\right)$, and $\bar{Z}_{hj}:=\frac{1}{n}\sum_{i=1}^n \bar{Z}_{hij}$. If $K$ is chosen  as in section \ref{KC}, and under \Cref{AfXmin,Afreg,Afsparse},
\begin{enumerate}
\item 
Under \Cref{AfXmin,Afreg,Afsparse},
for $j\notin\mathcal{R}$:
\begin{equation*}
\E{ \bar{Z}_{hj}}=0.
\end{equation*}
whereas,  for $j\in\mathcal{R}$, 
for $n$ large enough, 
\begin{equation}\label{e2.lmZbar1}
\frac12 \CEZ h_j^{p-1}
\leq \left\vert\E{\bar{Z}_{hj}}\right\vert
\leq \frac32 \CEZ h_j^{p-1},
\end{equation}
where $\CEZ:=\left\vert\tfrac{ \int_{\R}t^p K(t)dt}{(p-1)!}\partial_j^pf(w)\right\vert$.\\
 Besides, let $\CEZA
 :=\Vert f\Vert_{\infty,\UUn} \Vert J\Vert_1 \Vert K\Vert_1^{d-1}$. Then :
 \begin{equation}\label{e2.lmZbar2}
   \E{\vert \bar{Z}_{h1j}\vert}
 \leq \CEZA h_j^{-1}.
\end{equation}  
\item Let $\BZ{hj}:=\lbrace \vert \bar{Z}_{hj}-\E{\bar{Z}_{hj}}\vert\leq\frac12\seuil{hj} \rbrace$.
Under \Cref{AfXmin,Afreg,Afsparse},
if the bandwidth satisfies:
\begin{enumerate}
\item[$\text{Cond}_{\bar{Z}}(h)$:] $ \prod\limits_{k=1}^d h_k
\geq   \condZbar\frac{(\log n)^a}{n}$, with $\condZbar:=\frac{4 \Vert J\Vert_\infty^2\Vert K\Vert_\infty^{2(d-1)}}{ 3^2 \Vert f \Vert_{\infty,\UUn} \Vert J\Vert_2^2 \Vert K\Vert_2^{2(d-1)}} $,
\end{enumerate} 
then: 
$\PP{\BZ{hj}^c}\leq 2 e^{-\gZ},$
with 
$\gZ:=\frac{\delta }{\Vert f\Vert_{\infty,\UUn}} (\log n)^a$.
\item Let $\BZA{hj}:=\lbrace\vert \frac1n \sum\limits_{i=1}^n \vert\bar{Z}_{hij}\vert -\E{\vert\bar{Z}_{h1j}\vert}\vert\leq \CEZA h_j^{-1} \rbrace$. Then, under \Cref{AfXmin,Afreg,Afsparse}:
$$
\PP{\BZA{hj}^c}\leq 2 e^{-\CgZA n\prod_{k=1}^d h_k},
$$
with $\CgZA:= \min\left(\frac{\delta\CEZA^2}{4\Vert f\Vert_{\infty,\UUn} \Vert J\Vert_2^2 \Vert K\Vert_2^{2(d-1)}} ; \frac{3\delta\CEZA}{4\Vert K\Vert_\infty^{d-1}\Vert J\Vert_\infty}\right)$.
\end{enumerate}
\end{lem}


\begin{lem}\label{lmD}
For any $h\in(0,h_0]^d$ and any component $j= 1:d$, we denote $\DZ{hj}:=Z_{hj}-\bar{Z}_{hj}$ and $\Delta_h:= \f{h}-\fbar{h}$.
Under \Cref{Afsparse,Afreg,AfXmin}, if the conditions on $\fX$ are satisfied (see section \ref{sectionEstimatorChoice}), then, 
\begin{enumerate}
\item for $\CMDZ:=\frac{2 \CEZA M_X}{\Cl}$:
\begin{equation*}
\ind_{\BZA{hj}\cap\An}\left\vert\DZ{hj}\right\vert 
\leq  \frac{\CMDZ}{ (\log n)^{\frac{a}2}} \seuil{hj}
\end{equation*}
\item for $\CMD:=\frac{2\CEbar M_X}{\Cs}$:
 \begin{equation*}
\ind_{\An\cap\BA{h}} \left\vert\Delta_{h}\right\vert
\leq  \frac{\CMD}{ (\log n)^{\frac{a}{2}}} \sigma_h.
\end{equation*}
\end{enumerate}
\end{lem}

\noindent We introduce the notation $h^{(t)}$, $t\in\mathbb{N}$, the state of the bandwidth at iteration $t$ if $\hat{h}=h$. In particular for a fixed $t\in\lbrace0,\dots,\lfloor \preA\rfloor\rbrace$, $h^{(t)}$ is identical for any $h\in\HGP$. Then we consider the event:
 $$\EZ :=\An 
\cap \bigcap\limits_{j\notin\rel} \left\lbrace \BZ{h^{(0)}j} \cap \BZA{h^{(0)}j}\right\rbrace 
\cap  \bigcap\limits_{j\in\rel}\left[
\bigcap\limits_{h\in\HGP} \left\lbrace \BZ{hj} \cap \BZA{hj}\right\rbrace 
\cap \bigcap\limits_{t=0}^{\lfloor\preA\rfloor} \left\lbrace \BZ{h^{(t)}j} \cap \BZA{h^{(t)}j}\right\rbrace
\right],$$
where  we denote $\An:=\left\lbrace\sup\limits_{u\in\Un}\left\vert \frac{f_X(u)-\fX(u)}{\fX(u)}\right\vert \leq M_X \frac{(\log \nX)^b}{\nX^{\alpha}} \right\rbrace$.\\

\begin{prop}[\CDRodeo behaviour]\label{propRod}
Under \Cref{AfXmin,Afreg,Afsparse}, on $\EZ$, $\hat{h}\in\HGP$.\\
In other words, when $\EZ$ happens:
\begin{enumerate}
\item  non relevant components are deactivated during the iteration $0$; 
\item at the end of the iteration $\lfloor\preA\rfloor$, the active components are exactly the relevant ones; 
\item \CDRodeo{} stops at last at the iteration $\lfloor\postA\rfloor$.
\end{enumerate}
Moreover, for any $q>0$:
$$\PP{\EZ^c}=o(n^{-q}).$$ 
\end{prop}
\noindent The following lemma give a technical result to canonically obtain an upper bound of the bias of a kernel estimator.
Let us denote  $\cdot$  the multiplication terms by terms of two vectors. 
\begin{lem}\label{lmKernelBiasUpperBound}
Let $u\in\R^{d'}$ and a bandwidth $h\in\left(\R_+^*\right)^{d'}$. 
For $j=1:d'$, let $K:\R\rightarrow\R$ be a continuous function with compact support and with at least $p-1$ zero moments, \textit{ie}: for $l=1:(p-1)$, $$\int_{\R} K(t)t^ldt=0.$$ 
 Let $f_0:\R^{d'}\rightarrow\R$ a function of class $\mathcal{C}^p$ on $\mathcal{U}_h(u):=\left\lbrace u'\in\mathbb{R}^{d'}: \forall j=1:d', u'_j=u_j-h_jz_j, \text{with } z_j\in\supp(K_j)\right\rbrace$.
Then:
\begin{equation}
\int_{\R^{d'}} \left(\prod\limits_{j=1}^{d'} h_j^{-1}K\left(\tfrac{u_j-u'_j}{h_j}\right)\right) f_0(u') du' - f_0(u) \int_{\R^{d'}} \left(\prod\limits_{j=1}^{d'} K\left(z_j\right)\right) dz
=
\sum\limits_{k=1}^{d'} 
(\text{I}_k 
+\text{II}_k),
\end{equation}
where
$$\text{I}_k:=\int_{z\in\mathbb{R}^{d'}} \left(\prod\limits_{j=1}^{d'} K(z_{j})\right) \rho_k dz,$$
with the notations $\rho_k:=\rho_k(z,h,u)
=(-h_kz_k)^p\int\limits_{0\leq t_p\leq \dots\leq t_1\leq 1} 
\left(\partial_k^pf_0(\overline{z}_{k-1}-t_ph_kz_ke_k)-\partial_k^pf_0(\overline{z}_{k-1})\right)dt_{1:p},$
and  
$\overline{z}_{k-1}:=u-\sum\limits_{j=1}^{k-1} h_{j}z_{j}e_{j}$ (where $\lbrace e_{j}\rbrace_{j=1}^{d'}$ is the canonical basis of $\R^{d'}$),
and
$$\text{II}_k:=(-h_k)^p \int_{t\in\mathbb{R}} \frac{t^p}{p!} K(t) dt\int_{z_{-k}\in\R^{d'-1}}\partial_k^p f_0(\overline{z}_{k-1}) \left(\prod\limits_{j\neq k} K(z_{j})\right)
dz_{-k}.$$

\end{lem}
\noindent
Finally, we recall (without proof) the classical Bernstein's Inequality and Taylor's theorem with integral remainder.
\begin{lem}[Bernstein's inequality]\label{lmBernstein}
Let $U_1,\dots,U_n$  be independent random variables almost surely uniformly bounded by a positive constant $c>0$ and such that for $i=1,\dots,n$, $\mathbb{E}[U_i^2]\leq v$ . Then for any $\lambda>0$,
$$ \mathbb{P}\left(\left|\frac{1}{n}\sum_{i=1}^n U_i-\mathbb{E}[U_i]\right|\geq \lambda\right)\leq
2\exp\left(-\min\left(\frac{n\lambda^2}{4v},\frac{3n\lambda}{4c}\right)\right).$$
\end{lem}
\noindent
Note that this version is a simple consequence of Birgé and Massart (p.366 of \citep{BM98}).

\begin{lem}[Taylor's theorem] \label{lmTaylor}
Let $g:[0,1]\rightarrow\mathbb{R}$ be a function of class $\mathcal{C}^q$. Then we have:
\begin{equation*}
g(1)-g(0)=\sum\limits_{l=1}^q \frac{g^{(l)}(0)}{l!} +\int_{t_1=0}^1 \int_{t_2=0}^{t_1}\dots \int_{t_q=0}^{t_{q-1}} (g^{(q)}(t_q)-g^{(q)}(0)) dt_q dt_{q-1}\dots dt_1
\end{equation*}
\end{lem}

\subsection{Proofs of \Cref{propfXtilde}, \Cref{thmRodeo} and \Cref{cor}}\label{ssectProofMainResults}
\subsubsection{Proof of \Cref{propfXtilde}}\label{sectProoffXtilde}
We construct $\fX$ in two steps: 
 we first construct an estimator $\fXK$ which satisfies   
\begin{equation} 
 \PP{\Vert f_X-\fXK\Vert_{\infty,\Un} > M_X\frac{(\log n)^{\frac34}}{n^{\frac12}} }\leq  \exp(-(\log n)^{\frac54}),\label{eCond1fXK}
 \end{equation}
 then we show that if we set $\fX\equiv \fXK\vee(\log n)^{-\frac14}$, $\fX$ satisfies Conditions \ref{fXtildemin} and \ref{fXtildeAccuracy} for $n$ large enough.\\
  
\noindent We take $\fXK$ as the kernel density estimator defined in\eqref{eDef fXK}, with a kernel $\KX:\R\rightarrow\R$ that is compactly supported, of class $\mathcal{C}^1$, of order $p_X\geq \frac{d_1}{2(c-1)}$. 
 and  a bandwidth $\hX\in\R_+^*$ specified later.
\noindent
Let us control the bias $\left\Vert\E{\fXK}- f_X\right\Vert_{\infty,\Un}$. 
We define $p'_X=\min(p',p_X+1)$. In particular, $f_X$ is of class $\mathcal{C}^{p'_X}$ and $\KX$ has $p'_X-1$ zero moments.\\
Therefore we can apply \Cref{lmfXK} : 
\begin{align*}
\left\Vert \E{\fXK}- f_X\right\Vert_{\infty,\Un}
&\leq \CbiasX' h_X^{p'_X},
\end{align*}
where $\CbiasX':=\frac{\Vert\KX\Vert_1^{d_1-1}\Vert\cdot^{p'_X}\KX(\cdot)\Vert_1}{p'_X!} d_1\max\limits_{k=1:d_1}\Vert \partial_k^{p'_X}f_X\Vert_{\infty,\Up}$.\\
Therefore, since
\begin{align*}
\left\Vert \fXK-f_X \right\Vert_{\infty,\Un}
&\leq \left\Vert \fXK-\E{\fXK} \right\Vert_{\infty,\Un} + \left\Vert \E{\fXK}-f_X \right\Vert_{\infty,\Un}
\\
&\leq \left\Vert \fXK-\E{\fXK} \right\Vert_{\infty,\Un} + \CbiasX' h_X^{p'_X},
\end{align*}
we have for any threshold $\lambda$:
\begin{align}
\PP{\left\Vert \fXK-f_X \right\Vert_{\infty,\Un}\geq \lambda}
&\leq \PP{\left\Vert \fXK-\E{\fXK} \right\Vert_{\infty,\Un}\geq \lambda-\CbiasX' h_X^{p'_X}}. \label{eProbSansBias}
\end{align}
Therefore, we have reduced the problem to a local concentration inequality of $\fXK$ in sup norm.
In order to move from a supremum on $\Un$ to a maximum on a finite set of elements of $\Un$, let us construct a $\epsilon$-net of $\Un$. 
We denote $A>0$ such that: 
$$\supp(\KX)\cup \supp(K)\subset \left[-\tfrac{A}2,\tfrac{A}2\right].$$
 We set $N(\epsilon)$ the smallest integer such that 
 $ \epsilon N(\epsilon)\geq \frac{A}{\log n},$ 
 \textit{i.e.}: 
 $$N(\epsilon):=\left\lceil\frac{A}{\epsilon\log n}\right\rceil, $$
  then we introduce the notation $u_{(l)}\in\Un$, for a multi-index $l\in\left(1:N(\epsilon)\right)^{d_1}$ defined, such that the $j^{\text{th}}$ component of $u_{(l)}$ is:
 $$u_{(l)j}:=x_j-\frac{A}{2\log n}+(2l_j-1)\frac{\epsilon}2.$$
 Then 
$\lbrace u_{(l)}: l\in(1:N(\epsilon))^{d_1}\rbrace$ is a $\epsilon$-net of $\Un$, in the meaning that for any $u\in\Un$, there exists $l\in\lbrace1,\dots, N(\epsilon) \rbrace^{d_1}$ such that $\Vert u-u_{(l)}\Vert_{\infty}:=\max\limits_{k=1:d_1}\vert u_k- u_{(l)k}\vert\leq \epsilon$.\\
Therefore to obtain the desired concentration inequality, we only need to obtain the concentration inequality for each point of $\lbrace u_{(l)}: l\in(1:N(\epsilon))^{d_1}\rbrace$ and to control the following supremum $$\sup\limits_{u\in\Un}\min\limits_{l\in(1:N(\epsilon))^{d_1}} \left\vert \fXK(u) -\E{\fXK(u) }- \fXK(u_{(l)})+\E{\fXK(u_{(l)}) } \right\vert.$$ 
For this purpose, we obtain (from Taylor's Inequality): for any $u,v\in\R^{d_1}$,
$$\left\vert\prod\limits_{k=1}^{d_1} \KX(u_k) - \prod\limits_{k=1}^{d_1} \KX(v_k) \right\vert 
\leq d_1 \Vert \KX'\Vert_{\infty} \Vert \KX\Vert_{\infty}^{d_1-1}\Vert u-v\Vert_{\infty}.$$

Therefore, for any $u,v\in\Un$:
\begin{align*}
\left\vert \fXK(u) - \fXK(v) \right\vert 
&\leq \frac{1}{\nX .\hX^{d_1}}\sum\limits_{i=1}^{\nX} \left\vert\prod\limits_{k=1}^{d_1} \KX(\tfrac{u_k-\X_{ik}}{\hX}) - \prod\limits_{k=1}^{d_1} \KX(\tfrac{u_k-\X_{ik}}{\hX}) \right\vert
\\
&\leq \frac{d_1}{\hX^{d_1+1}} \Vert \KX'\Vert_{\infty} \Vert \KX\Vert_{\infty}^{d_1-1}\Vert u-v\Vert_{\infty}.
\end{align*}
Since $\lbrace u_{(l)}: l\in(1:N(\epsilon))^{d_1}\rbrace$ is a $\epsilon$-net of $\Un$:
\begin{align*}
\sup\limits_{u\in\Un}\min\limits_{l\in(1:N(\epsilon))^{d_1}} \left\vert \fXK(u) - \fXK(u_{(l)}) \right\vert
&\leq \frac{d_1}{\hX^{d_1+1}} \Vert \KX'\Vert_{\infty} \Vert \KX\Vert_{\infty}^{d_1-1}\epsilon.
\end{align*}
Thus:
\begin{align*}
\sup\limits_{u\in\Un}\min\limits_{l\in(1:N(\epsilon))^{d_1}} \left\vert \E{\fXK(u)} - \E{\fXK(u_{(l)})} \right\vert
&\leq d_1 \Vert \KX'\Vert_{\infty} \Vert \KX\Vert_{\infty}^{d_1-1}\frac{\epsilon}{\hX^{d_1+1}}.
\end{align*}
And so:
\begin{align*}
\sup\limits_{u\in\Un}\min\limits_{l\in(1:N(\epsilon))^{d_1}} \left\vert \fXK(u)-\E{\fXK(u)} - \fXK(u_{(l)})+ \E{\fXK(u_{(l)})} \right\vert
&\leq 2d_1 \Vert \KX'\Vert_{\infty} \Vert \KX\Vert_{\infty}^{d_1-1}\frac{\epsilon}{\hX^{d_1+1}}.
\end{align*}
We denote $\text{C}_{\text{diff}}:=2d_1 \Vert \KX'\Vert_{\infty} \Vert \KX\Vert_{\infty}^{d_1-1}$.
Then:
\begin{align*}
\left\Vert \fXK -\E{\fXK}\right\Vert_{\infty,\Un} 
&\leq \max\limits_{l\in(1:N(\epsilon))^{d_1}} \left\vert \fXK(u_{(l)}) -\E{\fXK(u_{(l)})}\right\vert
\\
&\qquad + \sup\limits_{u\in\Un}\min\limits_{l\in(1:N(\epsilon))^{d_1}} \left\vert \fXK(u)-\E{\fXK(u)} - \fXK(u_{(l)})+ \E{\fXK(u_{(l)})} \right\vert
\\
&\leq \max\limits_{l\in(1:N(\epsilon))^{d_1}} \left\vert \fXK(u_{(l)}) -\E{\fXK(u_{(l)})}\right\vert
+\text{C}_{\text{diff}}\frac{\epsilon}{\hX^{d_1+1}}.
\end{align*}
Then the inequality \eqref{eProbSansBias} becomes: for any threshold $\lambda$,
\begin{align}
\PP{\left\Vert \fXK-f_X \right\Vert_{\infty,\Un}\geq \lambda}
&\leq \PP{\left\Vert \fXK-\E{\fXK} \right\Vert_{\infty,\Un}\geq \lambda-\CbiasX' h_X^{p'_X}} \nonumber
\\
&\leq \PP{\max\limits_{l\in(1:N(\epsilon))^{d_1}}\left\vert \fXK(u_{(l)})-\E{\fXK(u_{(l)})} \right\vert \geq \lambda-\CbiasX' h_X^{p'_X}-\text{C}_{\text{diff}}\frac{\epsilon}{\hX^{d_1+1}}}\nonumber
\\
&\leq  N(\epsilon)^{d_1} \max\limits_{l\in(1:N(\epsilon))^{d_1}}\PP{\left\vert \fXK(u_{(l)})-\E{\fXK(u_{(l)})} \right\vert \geq \lambda-\CbiasX' h_X^{p'_X}-\text{C}_{\text{diff}}\frac{\epsilon}{\hX^{d_1+1}}}
 \label{eProbSansBiasSansSup}
\end{align}
We want to apply 2. of \Cref{lmfXK}. 
Therefore we fix the following settings:
\begin{itemize}
\item $\hX:= \nX^{-\frac{c-1}{c.d_1}}$
\item $\lambda:=2\lambda_X$, where $\lambda_X$ is the threshold in 2. of \Cref{lmfXK};
\item $\epsilon:=\hX^{1+\frac{d_1}{2}}\nX^{-\frac12}$.
\end{itemize}
For short, we denote $\text{C}_{\lambda X}:=2\Vert \KX\Vert_2^{d_1}\Vert f_X\Vert_{\infty,\Up}^{\frac12}$, so:
\begin{align*}
\lambda_X
=\sqrt{\tfrac{4\Vert \KX\Vert_2^{2d_1}\Vert f_X\Vert_{\infty,\Up}}{\hX^{d_1}\nX}(\log n)^{\frac32}}
= \text{C}_{\lambda X} (\log n)^{\frac34}\hX^{-\frac{d_1}2}\nX^{-\frac12}
= \text{C}_{\lambda X} (\log n)^{\frac34}\nX^{-\frac1{2c}}. 
\end{align*}
In particular, since we take $p_X\geq \frac{d_1}{2(c-1)}$ and we assume $p'\geq \frac{d_1}{2(c-1)}$, then $p'_X=\min(p',p_X)\geq \frac{d_1}{2(c-1)}$. Hence we obtain for $n$ large enough:
\begin{align*}
\CbiasX' h_X^{p'_X} 
&= \CbiasX' 
\nX^{-\frac{p'_X(c-1)}{c.d_1}}
\\
&\leq \CbiasX'    \nX^{-\frac{1}{2c}}
\\
&\leq \frac12 \lambda_X 
=  \frac{\text{C}_{\lambda X}}{2} (\log n)^{\frac34}\nX^{-\frac1{2c}}.
\end{align*}
and also, since $c>1$:
\begin{align*} 
\text{C}_{\text{diff}}\frac{\epsilon}{\hX^{d_1+1}}
&= \text{C}_{\text{diff}} \hX^{-\frac{d_1}2}\nX^{-\frac12}
= \text{C}_{\text{diff}}\ \nX^{-\frac{1}{2c}}
\\
&\leq \frac12 \lambda_X
= \frac{\text{C}_{\lambda X}}{2} (\log n)^{\frac34}\nX^{-\frac1{2c}}.
\end{align*}
Thus:
$$\lambda-\CbiasX' h_X^{p'_X}-\text{C}_{\text{diff}}\frac{\epsilon}{\hX^{d_1+1}}\geq \lambda_X,$$ and the inequality \eqref{eProbSansBiasSansSup}  becomes:
\begin{align}\label{eProbSansConcentration}
\PP{\left\Vert \fXK-f_X \right\Vert_{\infty,\Un}\geq \lambda}
&\leq  N(\epsilon)^{d_1} \max\limits_{l\in(1:N(\epsilon))^{d_1}}\PP{\left\vert \fXK(u_{(l)})-\E{\fXK(u_{(l)})} \right\vert \geq \lambda_X}
\end{align}
We verify that $\text{Cond}_X(\hX)$ is satisfied for $n$ large enough:
\begin{align*}
 \hX^{d_1}
 &= \nX^{-\frac{c-1}{c}}
 \\
 &\geq \frac{4\Vert \KX\Vert_{\infty}^{2d_1}}{9\Vert \KX\Vert_2^{2d_1}\Vert f_X\Vert_{\infty,\Up}} \frac{(\log n)^{\frac32}}{\nX}.
\end{align*}
Then we can apply 2. of \Cref{lmfXK},
\begin{align*}
\PP{\left\vert\fXK(u_{(l)})-\E{\fXK(u_{(l)})}\right\vert>\lambda_X}
\leq 2 \exp\left(- (\log n)^{\frac32}\right).
\end{align*}
Thus the inequality \eqref{eProbSansConcentration} becomes:
\begin{align}\label{eProbAvecNeps}
\PP{\left\Vert \fXK-f_X \right\Vert_{\infty,\Un}\geq \lambda}
&\leq  2N(\epsilon)^{d_1} \exp\left(- (\log n)^{\frac32}\right).
\end{align}
Let us control $ 2N(\epsilon)^{d_1}$:
 \begin{align*}
 2N(\epsilon)^{d_1}
 &=2\left\lceil\frac{A}{\epsilon\log n}\right\rceil^{d_1}
 \\
 &= 2\left\lceil\frac{A}{\hX^{1+\frac{d_1}{2}}\nX^{-\frac12}\log n}\right\rceil^{d_1}  
 \\
 &= o\left( \nX^{d_1+1}\right)
 \end{align*}
Then for $n$ large enough: 
 $$2N(\epsilon)^{d_1} \exp\left(- (\log n)^{\frac32}\right)\leq  \exp\left(- (\log n)^{\frac54}\right).$$
  Therefore:
 \begin{align*}
 \PP{\left\Vert \fXK-f_X \right\Vert_{\infty,\Un}\geq \lambda}
&\leq   \exp\left(- (\log n)^{\frac54}\right)
 \end{align*}
Since $\lambda=2\text{C}_{\lambda X} (\log n)^{\frac34}\nX^{-\frac1{2c}}$, we have obtained the desired concentration inequality \eqref{eCond1fXK} with $M_X=2\text{C}_{\lambda X}$.\\
~\\
Now we consider $\fX\equiv \fXK\vee (\log n)^{-\frac{1}{4}}$. By construction, $\fX$ satisfies Condition~\ref{fXtildemin}. 
Let us show it also satisfies Condition \ref{fXtildeAccuracy}, \textit{i.e.}:
$$\PP{\sup\limits_{u\in\Un}\left\vert \frac{f_X(u)-\fX(u)}{\fX(u)}\right\vert > M_{X} \frac{(\log n)^{\frac{d}2}}{n^{\frac12}}}\leq C_{X} \exp(-(\log n)^{\frac54}).$$
We write:
\begin{align*}
\PP{\sup\limits_{u\in\Un}\left\vert \frac{f_X(u)-\fX(u)}{\fX(u)}\right\vert > M_{X} \frac{(\log n)^{\frac{d}2}}{n^{\frac12}}}
&=
\PP{\exists u\in\Un, \left\vert f_X(u)-\fX(u)\right\vert > \fX(u) M_{X} \frac{(\log n)^{\frac{d}2}}{n^{\frac12}}}
\\
&\leq 
\PP{\exists u\in\Un, \left\vert f_X(u)-\fX(u)\right\vert > (\log n)^{-\frac{1}{4}} M_{X} \frac{(\log n)^{\frac{d}2}}{n^{\frac12}}}
\\
&\leq 
\PP{ \left\Vert f_X(u)-\fX(u)\right\Vert_{\infty,\Un} >  M_{X} \frac{(\log n)^{\frac{d}2-\frac14}}{n^{\frac12}}}
\end{align*}
Since $d=d_1+d_2\geq 2$,  $\frac{d}2-\frac14\geq \frac34$, we obtain from the previously proved concentration inequality \eqref{eCond1fXK}:
\begin{align*}
\PP{\sup\limits_{u\in\Un}\left\vert \frac{f_X(u)-\fX(u)}{\fX(u)}\right\vert > M_{X} \frac{(\log n)^{\frac{d}2}}{n^{\frac12}}}
&\leq  \PP{\left\Vert \fXK-f_X \right\Vert_{\infty,\Un}\geq   M_{X} \frac{(\log n)^{\frac34}}{n^{\frac12}}}
\\
&\leq   \exp\left(- (\log n)^{\frac54}\right).
\end{align*}

\subsubsection{Proof of \Cref{thmRodeo}}\label{sectionPrfThm}
We introduce $\Ef:=\bigcap\limits_{h\in\HGP} \left(\B{h}\cap\BA{h}\right)$ and denote $\mathcal{E}:=\EZ\cap\Ef$. 
On $\mathcal{E}$, $\hat{h}$ belongs to $\HGP$ (cf \Cref{propRod}). Thus:
\begin{equation}\label{eDecomp hath in HGP}
\ind_{\mathcal{E}}\left(\f{\hat{h}}-f(w)\right)
=\ind_{\mathcal{E}}\sum\limits_{h\in\HGP} \ind_{\hat{h}=h} \left(\f{h}-f(w)\right).
\end{equation} 
For any $h\in\HGP$, we denote $\Delta_h:=\f{h}-\fbar{h}$ and $\overline{B}_h:=\E{\fbar{h}}-f(w)$, and we decompose the loss as follows:
\begin{align}
\left\vert \f{h}-f(w) \right\vert
\leq \left\vert \Delta_h \right\vert
+ \left\vert \fbar{h}- \E{\fbar{h}} \right\vert
+ \left\vert \overline{B}_h \right\vert \label{eDecompfbar}.
\end{align}
Using \Cref{lmD},  since $\mathcal{E}\subset\An\cap \BA{h}$:
\begin{equation}\label{eMD}
\ind_{\mathcal{E}} \left\vert \Delta_h \right\vert
\leq \frac{\CMD}{(\log n)^{\frac{a}{2}}} \sigma_h.
\end{equation}
Moreover, by \Cref{lmfbar}, since $\mathcal{E}\subset\An\cap \B{h}$:
 \begin{align}
 \left\vert \fbar{h}- \E{\fbar{h}} \right\vert
 &\leq \sigma_h
 \label{eOnBbarh}\\
 &=\Cs\sqrt{ \frac{(\log n)^a}{ n \prod\limits_{k=1}^d h_k}}
 \nonumber \\
 &\leq \Cs \sqrt{ \frac{(\log n)^a}{ n h_0^d \beta^{r(\postA-\preA)+r\preA}}}
  \nonumber\\
 &=\Cs \CT^{\frac{-r}{2(2p+1)}}\Ctau^{\frac{-r}{2(2p+r)}}(\log n)^{\frac{p(a+d-r)}{2p+r}} n^{-\frac{p}{2p+r}}.\label{eMsh}
 \end{align}
And, also: 
\begin{align}
\vert \overline{B}_h\vert
 \leq \Cbiasbar \sum\limits_{k\in\mathcal{R}} h_k^p
 \leq r\Cbiasbar \beta^{p\preA} h_0^p
 = r\Cbiasbar \Ctau^{\frac{p}{2p+r}} (\log n)^{\frac{p(a+d-r)}{2p+r}} n^{-\frac{p}{2p+r}}.\label{eMbias}
\end{align}
%
%
To conclude, 
\begin{align*}
\ind_{\mathcal{E}}\vert\f{\hat{h}}-f(w)\vert
&\leq \ind_{\mathcal{E}}\sum\limits_{h\in\HGP} \ind_{\hat{h}=h} \left\vert\f{h}-f(w)\right\vert, 
\text{ by \eqref{eDecomp hath in HGP}}\nonumber
\\
& \leq \ind_{\mathcal{E}}\sum\limits_{h\in\HGP} \ind_{\hat{h}=h} \left(\left\vert \Delta_h \right\vert
+ \left\vert \fbar{h}- \E{\fbar{h}} \right\vert
+ \left\vert \overline{B}_h \right\vert\right), 
\text{ by \eqref{eDecompfbar}}\nonumber
\\
&\leq \ind_{\mathcal{E}}\sum\limits_{h\in\HGP} \ind_{\hat{h}=h} \left[\left(1+\frac{\CMD}{(\log n)^{\frac{a}{2}}}\right)\sigma_h
+ \left\vert \overline{B}_h \right\vert\right], \text{ by \eqref{eMD} and \eqref{eOnBbarh}}\nonumber
\\
&\leq \ind_{\mathcal{E}}\sum\limits_{h\in\HGP} \ind_{\hat{h}=h} \text{C} (\log n)^{\frac{p(a+d-r)}{2p+r}} n^{-\frac{p}{2p+r}},
\text{ by \eqref{eMsh} and \eqref{eMbias}}\nonumber
\\
&=\ind_{\mathcal{E}}\text{C} (\log n)^{\frac{p(a+d-r)}{2p+r}} n^{-\frac{p}{2p+r}},\nonumber
\end{align*}
 with for $n$ large enough (\textit{ie}: $\frac{\CMD}{(\log n)^{\frac{a}{2}}}\leq 1$), 
 $$\text{C}
:= r\Cbiasbar \Ctau^{\frac{p}{2p+r}} + 2\Cs \CT^{\frac{-r}{2(2p+1)}}\Ctau^{\frac{-r}{2(2p+r)}}.$$

\noindent It remains to give an upper bound on $\PP{\mathcal{E}^c}$. For any $q>0$: 
\begin{align*}
\PP{\mathcal{E}^c}
&\leq \PP{\EZ^c}+\PP{\Ef^c}\\
&\leq o(n^{-q}) + \sum\limits_{h\in\HGP} \left(\PP{\B{h}^c} + \PP{\BA{h}^c}\right), 
\text{ using \Cref{propRod}}\\
&\leq o(n^{-q}) + \sum\limits_{h\in\HGP} \left( 2e^{-(\log n)^a}+ 2 e^{-\CgfA n\prod_{k=1}^d h_k}\right), 
\end{align*}
using \Cref{lmfbar}, since for any $h\in\HGP$, Cond($h$) is satisfied.
Moreover: 
$$n\prod\limits_{k=1}^d h_k 
\geq n\beta^{r\postA}h_0^d
=\CT^{\frac{r}{2p+1}}\Ctau^{\frac{r}{2p+r}}(\log n)^{\frac{ra-2p(d-r)}{(2p+r)}} n^{\frac{2p}{2p+r}}
\geq \frac{(\log n)^a}{\CgfA},$$
for $n$ large enough.
Hence:
\begin{align*}
\PP{\mathcal{E}^c}
&\leq o(n^{-q}) + \vert\HGP\vert 4e^{-(\log n)^a}=o(n^{-q}),
\end{align*}
for $n$ large enough, since $\vert\HGP\vert=(\lceil\postA\rceil-\lfloor\preA\rfloor)^r==\left(\frac{1}{(2p+1)(\log(\frac{1}{\beta}))}\log(\frac{C_T}{C_\tau})+1\right)^r$ is finite.

\subsubsection{Proof of \Cref{cor}}
We consider the event $\mathcal{E}=\left\lbrace\left\vert \f{\hat{h}}-f(w)\right\vert 
\leq \text{C} (\log n)^{\frac{p}{2p+r}(d-r+a)} n^{-\frac{p}{2p+r}}\right\rbrace$ for which we proved in \Cref{thmRodeo}:
$$\PP{\mathcal{E}^c}=o(n^{-A}),$$
 for any $A>0$.
For short, we denote $ R_{h}:=\left\vert\f{h}-f(w)\right\vert$ for any bandwidth $h\in(\R_+^*)^d$.
Then we decompose $R_{\hat{h}}$ as follows:\begin{align*}
R_{\hat{h}}&=\mathds{1}_{\mathcal{E}}R_{\hat{h}}+\mathds{1}_{\mathcal{E}^c}R_{\hat{h}}.
\end{align*} 
By definition of $\mathcal{E}$, we immediately obtain:
\begin{align}\label{eCasProbable}
\mathds{1}_{\mathcal{E}}R_{\hat{h}} 
\leq \text{C} (\log n)^{\frac{p}{2p+r}(d-r+a)} n^{-\frac{p}{2p+r}}
\end{align}
For the second term, we first bound $\f{\hat{h}}$  a.s.
In \CDRodeo procedure, the loop stops when the current bandwidth becomes too small: $\prod\limits_{k=1}^d h_k<\frac{(\log n)}{ n}$. So the final bandwidth $\hat{h}$ satisfies:
$$\prod\limits_{k=1}^d \hat{h}_k\geq\frac{(\log n)}{ \beta^d n}.$$
Since $
\f{\hat{h}}=\dfrac{1}{n}\sum\limits_{i=1}^n \frac{1}{f_X(X_i)}\left(\prod\limits_{k=1}^d \hat{h}_k^{-1} K\left(\tfrac{w_j-W_{ij}}{\hat{h}_j}\right)\right)
$,
\begin{align*}
\left\vert \hat{f}_{\hat{h}}(w)\right\vert
&\leq \frac{\beta^d \Vert K\Vert_{\infty}^d}{\delta }\frac{n}{\log n}
\end{align*}
Hence: 
\begin{equation*}
R_{\hat{h}}\leq f(w)+ \frac{\beta^d \Vert K\Vert_{\infty}^d}{\delta }\frac{n}{(\log n)}
\end{equation*}
Therefore, for any $q>0$, using $(a+b)^q\leq 2^{q-1}(a^q+b^q)$:
\begin{align}
\E{\mathds{1}_{\mathcal{E}^c} (R_{\hat{h}})^q}
&\leq P(\mathcal{E}^c) 2^{q-1}\left(f(w)^q+ \Vert K\Vert_{\infty}^q\frac{n^q}{(\log n)^q}\right)\nonumber
 \\
& = o\left( n^{-A'}\right),\label{eCasNonProbable}
\end{align}
 for any $A'>0$ (since $P(\mathcal{E}^c)=o(n^{-A'+q})$).\\
We conclude by combining \eqref{eCasProbable} and \eqref{eCasNonProbable}:
\begin{equation}
\left(\E{\left\vert\f{\hat{h}}-f(w)\right\vert^q}\right)^{1/q}\leq \text{C}(\log n)^{\frac{p}{2p+r}(d-r+a)} n^{-\frac{p}{2p+r}}
+o\left(n^{-1}\right).
\end{equation}

\subsection{Proof of \Cref{propRod} and the lemmas}\label{ssectProofIntermediate}
\subsubsection{Proof of \Cref{propRod}}
First, note that the final state of the bandwidth determines exactly at which iteration each component has been deactivated: for a fixed bandwidth $h\in(\R_+^*)^d$,
if  $\hat{h}=h$, we denote $\lbrace\theta_k\rbrace_{k=1}^d$ such as for $k=1:d$, $h_k=h_0\beta^{\theta_k}$. In particular, $\theta_k$ is the iteration of deactivation of the component $k$.\\
We introduce the notation $h^{(t)}$, $t\in\mathbb{N}$, the state of the bandwidth at iteration $t$ if $\hat{h}=h$. 
It implies that $h^{(t)}$ is exactly defined by:
$h_k^{(t)}=\beta^{\theta_k\wedge t}h_0$ for $k=1:d$. \\
Notice that for a fixed $t\in\lbrace0,\dots,\lfloor \preA\rfloor\rbrace$, $h^{(t)}$ is identical for any $h\in\HGP$: by definition of $\HGP$, $h_j^{(t)}=h_0\beta^{t}$  if $j\in\rel$, else  $h_j^{(t)}=h_0$.
%
%
~\\
We recall the definition $$\EZ :=\An 
\cap \bigcap\limits_{j\notin\rel} \left\lbrace \BZ{h^{(0)}j} \cap \BZA{h^{(0)}j}\right\rbrace 
\cap \bigcap\limits_{j\in\rel}\left[
\bigcap\limits_{h\in\HGP} \left\lbrace \BZ{hj} \cap \BZA{hj}\right\rbrace 
\cap \bigcap\limits_{t=0}^{\lfloor\preA\rfloor} \left\lbrace \BZ{h^{(t)}j} \cap \BZA{h^{(t)}j}\right\rbrace
\right] .$$
For any component $j$ and any bandwidth $h$, we decompose $Z_{hj}$ as follows:
\begin{align}
\ind_{\EZ}Z_{hj}
&=\ind_{\EZ} \bar{Z}_{hj}
+ \ind_{\EZ}\DZ{hj}
\nonumber\\
&= \ind_{\EZ}  \E{\bar{Z}_{hj}}
+  \ind_{\EZ} (\bar{Z}_{hj}-\E{\bar{Z}_{hj}})
+ \ind_{\EZ}\DZ{hj}\label{eDecompZ}
\end{align}

\begin{enumerate}
\item Let us fix $j\notin\rel$ and $h=h^{(0)}=(h_0,\dots,h_0)$. Using 2. of \Cref{lmZbar}, $\E{\bar{Z}_{hj}}=0$. Therefore:
\begin{align*}
 \ind_{\EZ}\vert Z_{hj}\vert
&\leq   \ind_{\EZ} \left\vert\bar{Z}_{hj}-\E{\bar{Z}_{hj}}\right\vert
  + \ind_{\EZ}\left\vert \DZ{hj}\right\vert\\
 & \leq \frac12 \seuil{hj} +\ind_{\EZ}\left\vert \DZ{hj}\right\vert,
\end{align*}
using 2. of \Cref{lmZbar}, since $\EZ\subset \BZ{hj}$.
Now using 1. of \Cref{lmD}, since $\EZ\subset\BZA{hj} \cap \An$, we obtain:
\begin{align*}
 \ind_{\EZ}\vert Z_{hj}\vert
 & \leq \frac12 \seuil{hj} +\frac{\CMDZ}{ (\log n)^{\frac{a}2}} \seuil{hj}.
\end{align*}
Then for $n$ large enough (\textit{ie}: $(\log n)^{\frac{a}2}> 2\CMDZ$),  $\ind_{\EZ}\vert Z_{hj}\vert<\seuil{hj}$.
In other words, when $\EZ$ happens, all irrelevant components deactivate at the iteration $0$.

\item Let us show that $\EZ$ implies that the relevant components remain active until iteration $\lfloor\preA\rfloor+1$.\\
 It suffices to prove $\vert Z_{h^{(t)}j}\vert>\seuil{h^{(t)}j}$, for any $j\in\rel$ and any bandwidth $h^{(t)}$,  $t=0:\lfloor\preA\rfloor$.  (Indeed, by induction:  $(h_0,\dots,h_0)=h^{(0)}$, and since the irrelevant components deactivate at the iteration 0,  if the current bandwidth at the iteration $t$ is $h^{(t)}$ , then the fact that all the relevant components remain active for this bandwidth implies that the bandwidth at iteration $t+1$ is $h^{(t+1)}$).
 

 Let us fix $j\in\rel$, $t=0:\lfloor\preA\rfloor$ and we denote $h=h^{(t)}$. Using the decomposition \eqref{eDecompZ}, we obtain the following lower bound:
\begin{align*}
 \ind_{\EZ}\vert Z_{hj}\vert
&\geq \ind_{\EZ} \left( 
\left\vert \E{\bar{Z}_{hj}}\right\vert
-  \left\vert\bar{Z}_{hj}-\E{\bar{Z}_{hj}}\right\vert
-   \left\vert\DZ{hj}\right\vert
\right).
\end{align*}
Then, combining:
\begin{itemize}
\item[-]  $\left\vert \E{\bar{Z}_{hj}}\right\vert\geq \frac{\CEZ}{2} h_j^{p-1} $ \ (cf 1. of \Cref{lmZbar}), 
\item[-] $\left\vert\bar{Z}_{hj}-\E{\bar{Z}_{hj}}\right\vert\leq \frac12 \seuil{hj} $, \  since $\EZ\subset \BZ{hj} $ (cf 2. of \Cref{lmZbar}),
\item[-] $\left\vert\DZ{hj}\right\vert\leq \frac{\CMDZ}{ (\log n)^{\frac{a}2}} \seuil{hj}$, \ since 
$\EZ\subset \BZA{hj} \cap \An$ (cf 1. of \Cref{lmD}), 
\end{itemize}
   we obtain:
\begin{align*}
 \ind_{\EZ}\vert Z_{hj}\vert
&\geq \ind_{\EZ} \left( \frac{\CEZ}{2} h_j^{p-1}
-    \frac12 \seuil{hj} 
-  \frac{\CMDZ}{ (\log n)^{\frac{a}2}} \seuil{hj}
\right).
\end{align*}

Now let us show: $\ind_{\EZ}\vert Z_{hj}\vert
\geq \ind_{\EZ} \seuil{hj}.$\\
First, if $n$ is large enough (ie $(\log n)^{\frac{a}2}\geq 2 \CMDZ$), then 
$$ \frac{\CMDZ}{ (\log n)^{\frac{a}2}} \seuil{hj} \leq  \frac12 \seuil{hj} .  $$ 
Then it suffices to prove: 
$$\frac{\CEZ}{2} h_j^{p-1}\geq 2\seuil{hj},$$
\textit{i.e.}:
$$\qquad h_j^{2p}\prod\limits_{k=1}^d h_k\geq \frac{4^2\Cl^2}{\CEZ^2}\frac{(\log n)^a}n.$$
It is ensured for $t\leq \preA$, by definition of $\preA$ in \eqref{eDefpreA}:
\begin{align}
h_j^{2p}\prod\limits_{k=1}^d h_k
=\frac{\beta^{t(2p+r)}}{ (\log n)^{2p+d}}
\geq \frac{\beta^{\preA(2p+r)}}{(\log n)^{2p+d}}
=\frac{4^2\Cl^2}{\min\limits_{k\in\rel}\text{C}_{E\bar{Z},k}^2}\frac{(\log n)^a}n
\geq\frac{4^2\Cl^2}{\CEZ^2}\frac{(\log n)^a}n. \label{ebetataun}
\end{align}

Therefore, on $\EZ$, the component $j$ remains active until the iteration $\lfloor\preA\rfloor$.

\item Let us now prove that on $\EZ$, each relevant component $j$ deactivates at last  at iteration $\lceil \postA \rceil$. In particular, by definition of $\HGP$, $\hat{h}$ belongs to $\HGP$  on $\EZ$.\\ 
Assume $\EZ$ happens. \\
We fix $j\in\rel$. It suffices to prove that if $j$ is still active at iteration $\lceil \postA \rceil$, then on $\EZ$ happens, it deactivates at the end of this iteration. 
We assume $j$ is still active and we denote $h$ the state of the bandwidth at iteration $\lceil\postA\rceil$.

By the first point, for any $k\notin\rel$, $h_k=h_0$.\\ 
Given the second point, each relevant component $k$ was still active at the beginning of the iteration $\lfloor\preA\rfloor+1$,
 \textit{ie}: for any $k\in\rel$, $h_k\leq\beta^{\lfloor\preA\rfloor+1}h_0\leq\beta^{\preA}h_0$. 
 \\
 Moreover, since $j$ is still active, $h_j=\beta^{\lceil\postA\rceil}h_0$.
 Let us prove that: $\ind_{\EZ}\vert Z_{hj}\vert<\seuil{hj}$.
 Using the decomposition \eqref{eDecompZ}:
 \begin{align*}
 \ind_{\EZ}\vert Z_{hj}\vert
&\leq  \left\vert\E{\bar{Z}_{hj}}\right\vert +\ind_{\EZ} \left\vert\bar{Z}_{hj}-\E{\bar{Z}_{hj}}\right\vert
  + \ind_{\EZ}\left\vert \DZ{hj}\right\vert
 \end{align*}
Using the points 1. and 2. of \Cref{lmZbar} and 1. \Cref{lmD}, since $\EZ\subset \BZ{hj} \cap \BZA{hj} \cap \An$:
 \begin{align*}
 \ind_{\EZ}\vert Z_{hj}\vert
&\leq  2 \CEZ h_j^{p-1} +\frac12 \seuil{hj}
  + \frac{\CMDZ}{ (\log n)^{\frac{a}2}} \seuil{hj}\\
  &\leq  \seuil{hj} \left( \frac{2 \CEZ n^{\frac12} h_j^{p}\prod_{k=1}^d h_k^{\frac12}}{\Cl (\log n)^{\frac{a}2}} + \frac12 + \frac{\CMDZ}{ (\log n)^{\frac{a}2}} \right).
 \end{align*}
Given the specific form of $h$ : 
\begin{align*}
\frac{2 \CEZ n^{\frac12} h_j^{p}\prod_{k=1}^d h_k^{\frac12}}{\Cl (\log n)^{\frac{a}2}}
&\leq \frac{2 \CEZ n^{\frac12} h_0^{\frac{2p+d}2}\beta^{\frac{(2p+1)}2(\postA-\preA)} \beta^{\frac{(2p+r)\preA}{2}}}{\Cl(\log n)^{\frac{a}2}} \\
&= \sqrt{\frac{4^3 \CEZ^2 \beta^{(2p+1)(\postA-\preA)}}{ \min\limits_{k\in\rel}\text{C}_{E\bar{Z},k}^2 }},\text{ by definition of }\preA\\
&\leq \frac13, \text{  by definition of  } \postA.
\end{align*}
Moreover, for $n$ large enough: 
$$\frac{\CMDZ}{ (\log n)^{\frac{a}2}}< \frac16.$$
Therefore: $$\ind_{\EZ}\vert Z_{hj}\vert<\seuil{hj}.$$
 In other words, when $\EZ$ happens, any active component at iteration $\lceil\postA\rceil$ deactivates. 
\end{enumerate}
So we have proved that on $\EZ$, $\hat{h}\in\HGP$.\\
~\\
It remains to show that $\EZ$ holds with high probability.
\begin{align*}
\PP{\EZ^c }
&\leq \PP{\An^c} 
+ \sum\limits_{k=1}^d\left\lbrace\PP{ \BZ{h^{(0)}k}^c}+ \PP{ \BZA{h^{(0)}k}^c }\right\rbrace\\
&\quad + \sum\limits_{j\in\rel}\left[\sum\limits_{h\in\HGP} \left(\PP{ \BZ{hj}^c}+ \PP{ \BZA{hj}^c }\right)
+ \sum\limits_{t=1}^{\lfloor\preA\rfloor} \left( \PP{ \BZ{h^{(t)}j}^c}+\PP{ \BZA{h^{(t)}j}^c}\right)\right]
\end{align*}
By choice of $\fX$: $$\PP{\An^c} \leq C_X e^{-(\log n)^{\frac54}}.$$
We want to apply 2. and 3. of \Cref{lmZbar} for any $h\in\HGP$ and any $h^{(t)}$ with $t= 1:\lfloor\preA\rfloor$. These bandwidths satisfy:
$$\prod_{k=1}^d h^{(t)}_k
\geq\prod_{k=1}^d h_k
\geq h_0^d \beta^{r \lceil \postA\rceil}
\geq \CT^{\frac{r}{2p+1}}\Ctau^{\frac{r}{2p+r}} (\log n)^{\frac{ra-2p(d-r)}{2p+r}} n^{-\frac{r}{2p+r}},$$
which ensures that for $n$ large enough, $\text{Cond}_{\bar{Z}}(h^(t))$ and $\text{Cond}_{\bar{Z}}(h^{(t)})$ hold for any $h\in\HGP$ and any $t= 0:\lfloor\preA\rfloor$. Note in particular that $\HGP\subset\lbrace h^{(t),t= 0:\lceil\postA\rceil}$).\\
 Therefore, for any component $k=1:d$,
 $$\PP{ \BZ{h^{(0)}k}^c}\leq 2e^{-\gZ}$$ 
 and  for any $h\in\HGP$ and any $t= 0:\lceil\postA\rceil$ ,
\begin{align*}
\PP{ \BZA{h^{(t)}j}^c}
&\leq 2\exp\left(- \CgZA n\prod_{k=1}^dh_k\right)\\
&\leq 2\exp\left(- \CgZA \CT^{\frac{r}{2p+1}}\Ctau^{\frac{r}{2p+r}} (\log n)^{\frac{ra-2p(d-r)}{2p+r}}n^{\frac{2p}{2p+r}}\right)\\
&\leq 2e^{-\gZ}, \text{ for }n \text{ large enough.}
\end{align*} 
To conclude, note that $\vert\HGP\vert=(\lceil\postA\rceil-\lfloor\preA\rfloor)^r \leq (\postA-\preA+2)^r=\left(\frac{\log \left(\CT^{-1}\right)}{(2p+1)\log\frac1\beta}+2\right)^r$ is finite, so
 for any $q>0$:
 \begin{align*}
 \PP{\EZ^c }
&\leq C_X e^{-(\log n)^{\frac54}}+ 2(d+r\vert\HGP\vert+ r\preA) 2e^{-\gZ}\\
&=o(n^{-q}),
 \end{align*}
by definition $\gZ:=\frac{\delta }{\Vert f\Vert_{\infty,\UUn}} (\log n)^a$.

\subsubsection{Proof of \Cref{lmfXK}}
\begin{enumerate}
\item
We control the bias $\left\Vert\E{\fXK}- f_X\right\Vert_{\infty,\Un}$. We write for any $u\in\Un$:
\begin{align*}
\E{\fXK(u)}- f_X(u) 
&= \frac{1}{\hX^{d_1}}\int_{u'\in\R^{d_1}} \left(\prod\limits_{j=1}^{d_1} \KX\left(\tfrac{u_j-u'_j}{h_X}\right)\right) f_X(u')du' - f_X(u)\int_{\R^{d_1}}\left(\prod\limits_{j=1}^{d_1}\KX(z_j)\right)dz
\end{align*}
The kernel $\KX$ is of order $p_X$ and $f_X$ is assumed of class $\mathcal{C}^{p'}$ on $\Up$, with in particular $p'-1\leq p_X-1$,
then we can apply \Cref{lmKernelBiasUpperBound} with the settings  $u=u$, $d'=d_1$, $f_0=f_X$, $p=p'-1$, $K=K_X$ and for $j=1:d'$,  $h_k=\hX$. 
We obtain:
\begin{align}\label{eApplLemmeTechfXK}
\E{\fXK(u)}- f_X(u)
=\sum\limits_{k=1}^{d_1} (\text{I}_k +\text{II}_k).
\end{align}
with 
\begin{align*}
\text{I}_k&:=\int_{z\in\mathbb{R}^{d_1}} \left(\prod\limits_{k'=1}^{d_1} \KX(z_{k'})\right) \rho_k dz,
\\
\rho_k&:=\rho_k(z,h_X,u)\\
&=(-\hX z_k)^{p'-1}\int\limits_{0\leq t_{p'-1}\leq \dots\leq t_1\leq 1}  
\left(\partial_k^{p'-1}f_X(\overline{z}_{k-1}-t_{p'-1}h_Xz_ke_k)-\partial_k^{p'-1}f_X(\overline{z}_{k-1})\right) dt_{1:(p'-1)},
\\
\text{II}_k&:=(-h_X)^{p'-1} \int_{t\in\mathbb{R}} \frac{t^{p'-1}}{(p'-1)!} \KX(t) dt\int_{z_{-k}\in\R^{d_1-1}}\partial_k^{p'-1} f_X(\overline{z}_{k-1}) \left(\prod\limits_{k'\neq k} \KX(z_{k'})\right)
dz_{-k}.
\end{align*}
Let us control $\rho_k$. First we write:
\begin{align*}
\partial_k^{p'-1}f_X(\overline{z}_{k-1}-t_{p'-1}h_Xz_ke_k)-\partial_k^{p'-1}f_X(\overline{z}_{k-1})
&=-h_Xz_k\int_{t_{p'}=0}^1 \partial_k^{p'}f_X(\overline{z}_{k-1}-t_{p'}h_Xz_ke_k)dt_{p'}.
\end{align*}
Therefore:
\begin{align*}
\rho_k&
=(-\hX z_k)^{p'}\int\limits_{0\leq t_{p'}\leq \dots\leq t_1\leq 1}  
\partial_k^{p'}f_X(\overline{z}_{k-1}-t_{p'}h_Xz_ke_k)dt_{1:p'}.
\end{align*}
Hence:
\begin{align*}
 \vert \rho_k \vert
&\leq  \left\vert \hX z_k\right\vert^{p'}
\int\limits_{0\leq t_{p'}\leq \dots\leq t_1\leq 1} 
\left\vert\partial_k^{p'}f_X(\overline{z}_{k-1}-t_{p'}h_Xz_ke_k)\right\vert dt_{1:p'}
\\
&=  \frac{\vert z_k\vert^{p'}}{p'}\Vert \partial_k^{p'}f_X\Vert_{\infty,\Up}\hX^{p'}.
\end{align*}
Then:
\begin{align}
\left\vert\text{I}_k\right\vert
&\leq
\int_{z\in\mathbb{R}^{d_1}} \left\vert\prod\limits_{k'=1}^{d'} \KX(z_{k'})\right\vert \left\vert\rho_k\right\vert dz \nonumber
\\
&\leq \Vert \partial_k^{p'}f_X\Vert_{\infty,\Up}\hX^{p'} \int_{z\in\mathbb{R}^{d_1}}  \frac{\vert z_k\vert^{p'}}{p'} \left\vert\prod\limits_{k'=1}^{d'} \KX(z_{k'})\right\vert dz \nonumber
\\
&= \frac{\Vert\KX\Vert_1^{d_1-1}\Vert\cdot^{p'}\KX(\cdot)\Vert_1}{p'} \Vert \partial_k^{p'}f_X\Vert_{\infty,\Up} h_X^{p'} \label{eUpperBoundIkfXK}
\end{align}
Besides,  $\KX$ is of order $p_X$ and $p'-1<p_X$ and so:
\begin{align*}
\text{II}_k:=\frac{(-\hX)^{p'-1}}{(p'-1)!}\int_{t\in\mathbb{R}} t^{p'-1}\KX(t) dt  \int_{z_{-k}\in\R^{d_1-1}}\partial_k^{p'-1} f_X(\overline{z}_{k-1}) \left(\prod\limits_{k'\neq k} \KX(z_{k'})\right)dz_{-k} 
=0.
\end{align*}
Therefore the terms $\text{II}_k$ vanish in the equation \eqref{eApplLemmeTechfXK}, and with the upper bound of $\text{I}_k$ \eqref{eUpperBoundIkfXK}, we obtain:
\begin{align*}
\left\Vert \E{\fXK}- f_X\right\Vert_{\infty,\Un}
&= \sup\limits_{u\in\Un}  \left\vert\E{\fXK(u)}- f_X(u)\right\vert \nonumber
\\
&\leq  \sup\limits_{u\in\Un} \sum\limits_{k=1}^{d_1} \vert\text{I}_k\vert
\\
&\leq \frac{\Vert\KX\Vert_1^{d_1-1}\Vert\cdot^{p'}\KX(\cdot)\Vert_1}{p'!} h_X^{p'} \sum\limits_{k=1}^{d_1}  \Vert \partial_k^{p'}f_X\Vert_{\infty,\Up}
\\
&=\CbiasX h_X^{p'},
\end{align*}
with $\CbiasX:=\frac{\Vert\KX\Vert_1^{d_1-1}\Vert\cdot^{p'}\KX(\cdot)\Vert_1}{p'!} d_1\max\limits_{k=1:d_1}\Vert \partial_k^{p'}f_X\Vert_{\infty,\Up}$. 
\item
We apply Bernstein's inequality (see \Cref{lmBernstein}). We define for any $u\in\Un$ and any $i=1:\nX$: $$\fXKi{i}(u):=\frac{1}{\hX^{d_1}}\prod\limits_{j=1}^{d_1}\KX\left(\tfrac{u_j-\X_{ij}}{\hX}\right).$$ Then we control $\fXKi{1}$ a.s.: for any $u\in\Un$,
\begin{align*}
\left\vert\fXKi{1}(u)\right\vert
\leq  \text{M}_{\hX}:=\Vert \KX\Vert_{\infty}^{d_1} \hX^{-d_1}.
\end{align*} 
 and its variance:
 \begin{align*}
\V{\fXKi{1}(u)}
&\leq \E{(\fXKi{1})^2}
\\
&=\hX^{-2d_1}\int_{u'\in\R^{d_1}} \left(\prod\limits_{j=1}^{d_1}\KX\left(\tfrac{u_j-u'_{j}}{\hX}\right)\right)^2f_X(u')du'
\\
&=\hX^{-d_1}\int_{z\in\R^{d_1}} \left(\prod\limits_{j=1}^{d_1}\KX(z_j)\right)^2f_X(u-\hX z)du'
\\
&\leq \text{v}_{\hX}
\end{align*} 
with $\text{v}_{\hX}:=\text{C}_{\text{v}X}\hX^{-d_1}$ and 
$\text{C}_{\text{v}X}:=\Vert \KX\Vert_2^{2d_1}\Vert f_X\Vert_{\infty,\Up}$.\\
Then we apply \Cref{lmBernstein}: for any $\lambda>0$,
\begin{align*}
\PP{\left\vert\fXK(u)-\E{\fXK(u)}\right\vert>\lambda}
\leq 2 \exp\left(- \min\left(\frac{\nX \lambda^2}{4 \text{v}_{\hX}}, \frac{3\nX\lambda}{4\text{M}_{\hX}}\right)\right).
\end{align*}
We set $\lambda=\lambda_X:=\sqrt{\frac{4 \text{v}_{\hX}}{\nX}(\log n)^{\frac32}}
$ such that $(\log n)^{\frac32}=\frac{\nX \lambda^2}{4 \text{v}_{\hX}}$.
Then we compare the rates:
\begin{align*}
&\frac{\nX \lambda^2}{4 \text{v}_{\hX}}
\leq \frac{3\nX\lambda}{4\text{M}_{\hX}}
\\
\Longleftrightarrow\quad  &\lambda^2 \leq \frac{3^2 \text{C}_{\text{v}X}^2}{\Vert \KX\Vert_{\infty}^{2d_1}}
\\
\Longleftrightarrow\quad  &\hX^{d_1}
\geq \frac{4\Vert \KX\Vert_{\infty}^{2d_1}}{9\text{C}_{\text{v}X}} \frac{(\log n)^{\frac32}}{\nX}, 
\\
\Longleftrightarrow\quad  & \text{Cond}_X(\hX).
\end{align*}
\end{enumerate}

\subsubsection{Proof of \Cref{lmfbar}}
\begin{enumerate}
\item We recall the notation  $\cdot$ for  the multiplication terms by terms of two vectors. Then:
\begin{align*}
\vert\E{ \fbar{h1}} \vert
&\leq 
\E{\vert \fbar{h1}\vert} \\
&= \int_{u\in\mathbb{R}^d}  \left\vert \prod\limits_{k=1}^d \frac{K(h_k^{-1}(w_k-u_k))}{h_k} \right\vert f(u) du \\
&=\int_{z\in\mathbb{R}^d} \left\vert\prod\limits_{k=1}^d K(z_k)\right\vert f(w-h\cdot z) dz\\
&\leq \Vert f\Vert_{\infty,\UUn} \Vert K\Vert_{1}^d =:\CEbar
\end{align*}

Now let us give an upper bound on the bias of $\fbar{h}$:
\begin{align*}
\overline{B}_h
=\E{ \fbar{h1}} -f(w)
&=\int_{u\in\mathbb{R}^d} \left( \prod\limits_{k=1}^d \frac{K(h_k^{-1}(w_k-u_k))}{h_k} \right)f(u) du - f(w)\int_{\R^d}\prod\limits_{k'=1}^d K(z_{k'})dz,
\end{align*}
since $\int_{\R}K(t)dt=1$.
Then we apply the \Cref{lmKernelBiasUpperBound} with the settings  $d'=d$, $u=w$, $h=h$, $f_0=f$, $p=p$  and $K=K$. 
We obtain:
  \begin{align*}
\overline{B}_h
&=\sum\limits_{k=1}^d 
(\text{I}_k 
+\text{II}_k),
\end{align*}
where 
\begin{align*}
\text{I}_k
&:=\int_{z\in\mathbb{R}^d} \left(\prod\limits_{k'=1}^d K(z_{k'})\right) \rho_k dz,
\\
\rho_k
&:=(-h_kz_k)^p\int\limits_{0\leq t_p\leq \dots\leq t_1\leq 1} \left(\partial_k^pf(\overline{z}_{k-1}-t_ph_kz_ke_k)-\partial_k^pf(\overline{z}_{k-1})\right)dt_{1:p}\label{edefRhok},
\\
\text{II}_k
&:=(-h_k)^p \int_{t\in\mathbb{R}} \frac{t^p}{p!} K(t) dt\int_{z_{-k}\in\R^{d-1}}\partial_k^p f(\overline{z}_{k-1}) \left(\prod\limits_{k'\neq k} K(z_{k'})\right)
dz_{-k}.
\end{align*}
Notice that for $k\notin\rel$, $\partial_k^pf(u)=0$ for any  $u\in\Un$, thus $\text{I}_k$ and $\text{II}_k$ vanish. Therefore:
\begin{align*}
\overline{B}_h
&=\sum\limits_{k\in\mathcal{R}} 
(\text{I}_k 
+\text{II}_k),
\end{align*}
 Now let us give an equivalent of the bias.
 First, using \Cref{Afreg}, for any $k\in\rel$, we can define the modulus of continuity of $\partial_k^p f$ on $\UUn$ by:
\begin{equation*}
\Omega_{nk}
:= \sup\limits_{z,z'\in\UUn}\left\vert \partial_k^p f(z') - \partial_k^p f(z)  \right\vert
\end{equation*}
Then we decompose $\text{II}_k$ as follows:
$$\text{II}_k=\frac{(-h_k)^p \int_{t\in\mathbb{R}} t^p K(t) dt}{p!}
\partial_k^p f(w) + R_k,
 $$ with $R_k:=\frac{(-h_k)^p \int_{t\in\mathbb{R}} t^p K(t) dt}{p!}
\int_{z_{-k}\in\R^{d-1}}(\partial_k^p f(\overline{z}_{k-1})-\partial_k^p f(w) ) \left(\prod\limits_{k'\neq k} K(z_{k'})\right) dz_{-k}$ such that:
\begin{align}
\vert R_k\vert 
\leq h_k^p \left\vert\int_{t\in\mathbb{R}}  \frac{ t^p}{p!} K(t) dt\right\vert \Omega_{nk}\Vert K\Vert_1^{d-1}
\label{eMajRk}
\end{align} 
since 
$\left\vert\partial_k^p f(\overline{z}_{k-1})-\partial_k^p f(w)\right\vert
\leq \Omega_{nk}.$

It remains to bound $\text{I}_k$. From the definition of $\rho_k$ in \eqref{edefRhok}, we write:
\begin{align*}
\vert \rho_k\vert 
&\leq \vert h_kz_k\vert^{p}
\left\vert\int_{0\leq t_p\leq \dots\leq t_1\leq 1} \left[\partial_j^p f(\overline{z}_{k-1}-t_ph_kz_ke_k) -  \partial_j^p f(\overline{z}_{k-1})\right] dt_{1:p}\right\vert \\
&\leq  \vert h_kz_k\vert^{p} \frac{\Omega_{nk}}{p!}.
\end{align*}
Therefore: 
\begin{align}
\left\vert\text{I}_k\right\vert
&=\left\vert\int_{z\in\mathbb{R}^d} \left(\prod\limits_{k'=1}^d K(z_{k'})\right) \rho_k dz\right\vert \nonumber\\
&\leq   \frac{h_k^p}{p!}\Omega_{nk}\int_{z\in\mathbb{R}^d} \left\vert z_k^p\prod\limits_{k'=1}^d K(z_{k'})\right\vert  dz \nonumber\\
&\leq \Vert K\Vert_1^{d-1} \int_{t\in\R} \left\vert \frac{ t^p}{p!} K(t)\right\vert dt\times  h_k^{p} \Omega_{nk}\label{eMajIk}
\end{align}
 Since $\UUn\underset{n\rightarrow\infty}{\longrightarrow}  \lbrace w\rbrace$, by continuity of $\partial_k^p f$:
 $$\Omega_{nk}\underset{n\rightarrow\infty}{\longrightarrow} 0.$$
 Therefore for $n$ large enough, combining \eqref{eMajRk} and \eqref{eMajIk}:
 \begin{align*}
\left\vert\text{I}_k\right\vert+\vert R_k\vert \leq \frac{\vert \int_{t\in\R} t^p K(t) dt\vert}{p!}\max\limits_{k\in\rel} \vert \partial_k^p f(w)\vert \times h_k^p
 \end{align*}
Therefore, since: 
$$
 \overline{B}_h
 =\sum\limits_{k\in\mathcal{R}}(\text{II}_k+\text{I}_k)
= \sum\limits_{k\in\mathcal{R}} \left(\tfrac{(-h_k)^p \int_{t\in\mathbb{R}} t^p K(t) dt\  \partial_k^p f(w)}{p!}
 +R_k+\text{I}_k\right),
$$
we obtain:
 \begin{equation*}
 \left\vert \overline{B}_h\right\vert \leq \Cbiasbar \sum\limits_{k\in\mathcal{R}} h_k^p,
 \end{equation*}
with $\Cbiasbar:=  \frac{2\vert \int_{t\in\R} t^p K(t) dt\vert}{p!}\max\limits_{k\in\rel} \vert \partial_k^p f(w)\vert$.
\item  We want to apply Bernstein's inequality (cf \Cref{lmBernstein}) to $\fbar{h}$. We first obtain an almost sure upper bound:
\begin{align}
\vert\fbar{h1}\vert
&=\frac1{f_X(X_1)} \prod\limits_{k=1}^d \frac{\left\vert K\left(\frac{w_k-W_{1k}}{h_k}\right)\right\vert}{h_k}
\nonumber\\
&\leq 
\Mbar{h},\label{eUpperBoundfbaras}
\end{align}
where $\Mbar{h}:=\frac{\CMbar}{\prod_{k=1}^dh_k}$ with $\CMbar:=\frac{\Vert K \Vert^d_{\infty}}{\delta}$.\\
Then we control the variance:
\begin{align}
\V{\fbar{h1}}
&= \V{\frac1{f_X(X_1)}\prod\limits_{k=1}^d \frac{K\left(\frac{w_k-W_{1k}}{h_k}\right)}{h_k} }
\nonumber\\
&\leq \E{\left(\frac1{f_X(X_1)}\prod\limits_{k=1}^d \frac{K\left(\frac{w_k-W_{1k}}{h_k}\right)}{h_k} \right)^2}
\nonumber\\
&= \int_{u\in\mathbb{R}^d}  \left\lbrace\prod\limits_{k=1}^d \frac1{h_k^2}K\left(\frac{w_k-u_k}{h_k}\right)^2\right\rbrace \frac{f(u)}{f_X(u_{1:d_1})}du
\nonumber\\
&\leq   \frac{1}{\delta \prod\limits_{k=1}^d h_k}  \int_{z\in\mathbb{R}^d} \left\lbrace\prod\limits_{k=1}^d K(z_k)^2\right\rbrace f(w-Hz) dz
\nonumber\\
&\leq 
 \Vbar{h},\label{eUpperBoundVarfbar}
\end{align}
where $\Vbar{h}:=\frac{\Cs^2}{4\prod_{k=1}^d h_k}$.
 Therefore we obtain from Bernstein's inequality (cf \Cref{lmBernstein}): 
\begin{align*}
\PP{\B{h}^c}\leq 2 \exp\left( -\min\left(\frac{n\sigma_h^2}{4\Vbar{h}} , \frac{3n\sigma_h}{4\Mbar{h}}\right)\right)
\end{align*}
We compare the rates:
\begin{align*}
&\frac{n\sigma_h^2}{4\Vbar{h}} 
\leq \frac{3n\sigma_h}{4\Mbar{h}}
\\
\Longleftrightarrow\quad  &\Cs\sqrt{\frac{(\log n)^a}{n\prod_{k=1}^d h_k}}=\sigma_h 
\leq \frac{3\Vbar{h}}{\Mbar{h}}=\frac{3\Cs^2}{4\CMbar}
\\
\Longleftrightarrow\quad  &\prod_{k=1}^d h_k
\geq \frac{4\CMbar^2}{9\Cs^2} \frac{(\log n)^a}{n}
\\
\Longleftrightarrow\quad  & \text{Cond}(h).
\end{align*}
Therefore, if $\text{Cond}(h)$ is satisfied:
\begin{align*}
\PP{\B{h}^c}
\leq 2 e^{ -\frac{n\sigma_h^2}{4\Vbar{h}}}=2 e^{-(\log n)^a}.
\end{align*} 
\item  
We now apply Bernstein's inequality (cf \Cref{lmBernstein}) to $\frac1{n}\sum\limits_{i=1}^n \vert \fbar{hi}\vert $. From the upper bounds \eqref{eUpperBoundfbaras} and \eqref{eUpperBoundVarfbar}, we obtain:
 \begin{align*}
\PP{\BA{h}^c}
\leq 2 \exp\left( -\min\left(\frac{n\CEbar^2}{4\Vbar{h}} , \frac{3n\CEbar}{4\Mbar{h}}\right)\right).
\end{align*}
We calculate the rates: by definition of $\Vbar{h}$ and $\Mbar{h}$,
\begin{align*}
\frac{n\CEbar^2}{4\Vbar{h}}
=\frac{\CEbar^2}{\Cs^2}n\prod\limits_{k=1}^dh_k\\
\frac{3n\CEbar}{4\Mbar{h}}
= \frac{3\CEbar}{4\CMbar}n\prod\limits_{k=1}^dh_k
\end{align*}
Hence:
 \begin{align*}
\PP{\BA{h}^c}\leq 2 e^{-\CgfA n\prod_{k=1}^d h_k},
\end{align*}
with $\CgfA:= \min\left( \frac{\CEbar^2}{\Cs^2} ; \frac{3\CEbar}{4\CMbar} \right)$.
\end{enumerate}

\subsubsection{Proof of \Cref{lmZbar}}

\begin{enumerate}
\item 
First, we write $\bar{Z}_{hij}$ more explicitly: for any bandwidth $h$, any observation $i=1:n$ and any direction $j$,
\begin{align*}
\bar{Z}_{hij}
&=\frac{\partial}{\partial h_j}\left(\frac{K(\frac{w_j-W_{ij}}{h_j})}{h_j}\right)
\frac{\prod\limits_{k\neq j} {K}(\tfrac{w_k-W_{ik}}{h_k})}{f_X(X_i)\prod_{k\neq j} h_k}   \\
&=\frac{-\left(K(\frac{w_j-W_{ij}}{h_j})+\frac{w_j-W_{ij}}{h_j}K'(\frac{w_j-W_{ij}}{h_j})\right)
\prod\limits_{k\neq j} {K}(\tfrac{w_k-W_{ik}}{h_k})}{f_X(X_i)h_j\prod_{k=1}^d h_k}   \\
&= \frac{-J(\frac{w_j-W_{ij}}{h_j})\prod\limits_{k\neq j}{K}(\tfrac{w_k-W_{ik}}{h_k}) }{f_X(X_i)h_j\prod_{k=1}^d h_k}
\end{align*}
where we recall $J:\R\rightarrow\R$ is the function $t\mapsto tK'(t)+K(t)$. \\
Note then that the support of $J$ is included in the support of $K$, and by integration by part, we obtain for any $l\in\mathbb{N}$:
\begin{equation}
\int_{\R} t^l J(t)dt= \int_{\R} t^l (tK(t))'dt=-l \int_{\R} t^l K(t)dt\label{eEqalitéMomentJK}
\end{equation}
In particular, since $K$ is of order $p$, for $l=0:p-1$, $\int_{\R} t^l J(t)dt=0$ and $\int_{\R} t^p J(t)dt\neq 0$.\\
We recall the notation  $\cdot$ for  the multiplication terms by terms of two vectors.
Using \Cref{Afsparse}, if $j\notin\rel$, $f(w-h\cdot z)-f(\tilde{z}_{-j}) =0$ for any $z\in\R^{d}$. Thus we obtain:
\begin{align*}
 \E{\bar{Z}_{h1j}}
&=-\frac{1}{h_j\prod_{k=1}^d h_k}\int_{u\in\mathbb{R}^d }J(\tfrac{w_j-u_j}{h_j})\left(\prod\limits_{k\neq j}{K}(\tfrac{w_k-u_{k}}{h_k})\right) f(u)du
\\
&= -\frac{1}{h_j} \int_{z_j\in\R}J(z_j)dz_j\int_{z_{-j}\in\mathbb{R}^{d-1}} \left(\prod\limits_{k\neq j}{K}(z_k)\right) f(w-h\cdot z)dz_{-j}
=0
\end{align*}
Therefore $ \E{\bar{Z}_{h1j}}=0$ for $j\notin\rel$.\\
Now, we deal with the case $j\in\mathcal{R}$. Let us fix $j\in\rel$. We denote $\tilde{z}_{-j}:=w-(Hz)_{-j}=w-\sum\limits_{k\neq j}h_kz_ke_k$ (with $\lbrace e_k\rbrace_{k=1}^d$ the canonic basis of $\mathbb{R}^d$). Then we write:
\begin{align*}
 \E{\bar{Z}_{h1j}}
&=\frac{-1}{h_j\prod_{k=1}^d h_k}\int_{u_{-j}\in\mathbb{R}^{d-1} }\left(\prod\limits_{k\neq j}{K}(\tfrac{w_k-u_{k}}{h_k})\right) \left[
\int_{u_j\in\R}J(\tfrac{w_j-u_j}{h_j})du_j
- f(\tilde{z}_{-j})\int_{\R}J(z_j)dz_j
\right]du_{-j}.
\end{align*}
Then for fixed $\lbrace z_k\rbrace_{k\neq j}$, denoting $f_j:z_j\mapsto f(w-h\cdot z)$, we apply \Cref{lmKernelBiasUpperBound} with the settings  $d'=1$, $u=\tilde{z}_{-j}$, $h=h_j$, $f_0=f_j$, $p=p$, $K=J$, then 
\begin{align}
\E{\bar{Z}_{h1j}}
&=\frac{-1}{h_j\prod_{k=1}^d h_k}\int_{u_{-j}\in\mathbb{R}^{d-1} }\left(\prod\limits_{k\neq j}{K}(\tfrac{w_k-u_{k}}{h_k})\right) \left[
\text{I}_1 +\text{II}_1
\right]du_{-j} 
\nonumber \\
&= \widetilde{\text{I}}_j
+ \tilde{\text{II}}_j, \label{eEZhj2}
\end{align}
where
\begin{align}
\tilde{\text{I}}_j
&:=(-h_j)^{-1} \int_{z\in\mathbb{R}^d} \left(\prod\limits_{k\neq j} K(z_k)\right) J(z_j) \tilde{\rho}_j dz,\label{eDefITilde}\\
\text{with } \tilde{\rho}_j
&:=(-h_jz_j)^{p}\int\limits_{0\leq t_p\leq \dots\leq t_1\leq 1}\left(\partial_j^{p}f(\tilde{z}_{-j}-t_ph_jz_je_j)-\partial_j^{p}f(\tilde{z}_{-j})\right)dt_{1:p},\label{eDefRhoTilde}\\
\text{and } \widetilde{\text{II}}_j
&:=(-h_j)^{p-1} \int_{t\in\mathbb{R}} \frac{t^p}{p!} J(t) dt\int_{z_{-j}\in\R^{d-1}}\partial_j^p f(\tilde{z}_{j-1}) \left(\prod\limits_{k'\neq j} K(z_{k'})\right) dz_{-j}.\nonumber
\end{align}
Now let us determine an equivalent of  $\E{ \bar{Z}_{hj}} $.
For this purpose, let us introduce the modulus of continuity of $\partial_j^p f$ on $\UUn$ (which is well defined by \Cref{Afreg}):
\begin{equation*}
\Omega_{nj}
:= \sup\limits_{z,z'\in\UUn}\left\vert \partial_j^p f(z') - \partial_j^p f(z)  \right\vert
\end{equation*}
Then we write:
\begin{align}
\widetilde{\text{II}}_j
&= (-h_j)^{p-1}  \partial_j^p f(w)\int_{t\in\mathbb{R}} \frac{t^p}{p!} J(t) dt
+ \widetilde{R}_j,\label{eDecompIIjtilde}
\end{align}
with
\begin{equation*}
\widetilde{R}_j
 :=(-h_j)^{p-1} \int_{t\in\mathbb{R}} \frac{t^p}{p!} J(t) dt
\int_{z_{-j}\in\R^{d-1}}\left(\partial_j^p f(\tilde{z}_{-j})-\partial_j^p f(w) \right) \left(\prod\limits_{k\neq j} K(z_{k})\right) dz_{-j}.
\end{equation*}
In particular:
\begin{align}
\vert \widetilde{R}_j\vert 
&\leq h_j^{p-1} \left\vert\int_{t\in\mathbb{R}}  \frac{t^p}{p!} J(t) dt\right\vert
\int_{z_{-k}\in\R^{d-1}} \Omega_{nj} \prod\limits_{k\neq j} \vert K(z_{k})\vert dz_{-j} 
\nonumber\\
&= h_j^{p-1} \Omega_{nj} 
\left\vert\int_{t\in\mathbb{R}}  \frac{t^p}{p!} J(t) dt \right\vert 
\Vert K\Vert_1^{d-1}\label{eMajRtilde}
\end{align} 
Now let us bound $\tilde{\text{I}}_j$ defined in \eqref{eDefITilde}.
First, we bound $\tilde{\rho}_j$, defined in \eqref{eDefRhoTilde}:
\begin{align*}
\vert \tilde{\rho}_j\vert 
&=(h_j\vert z_j\vert)^{p}\left\vert \int\limits_{0\leq t_p\leq \dots\leq t_1\leq 1}\left(\partial_j^{p}f(\tilde{z}_{-j}-t_ph_jz_je_j)-\partial_j^{p}f(\tilde{z}_{-j})\right)dt_{1:p}\right\vert
\\
&\leq  h_j^p\vert z_j\vert^{p} \frac{\Omega_{nj}}{p!},
\end{align*}
which lead to:
\begin{align}
\left\vert \tilde{\text{I}}_j\right\vert
&=h_j^{-1} \left\vert \int_{z\in\mathbb{R}^d} \left(\prod\limits_{k\neq j} K(z_k)\right) J(z_j) \tilde{\rho}_j dz \right\vert
\nonumber\\
&\leq h_j^{p-1} \Omega_{nj}  \Vert K\Vert_1^{d-1}  \int_{z_j\in\R} \frac{ \vert z_j\vert^{p}}{p!} \vert J(z_j)\vert dz_j. \label{eMajItilde}
\end{align}

Therefore using \eqref{eDecompIIjtilde} then \eqref{eMajRtilde} and \eqref{eMajItilde}:
\begin{align*}
\left\vert\E{ \bar{Z}_{h1j}} \right\vert
&\leq \left\vert\widetilde{\text{II}}_j\right\vert
+ \left\vert\tilde{\text{I}}_j \right\vert 
\quad\leq \quad h_j^{p-1}  \left\vert\partial_j^p f(w)\int_{t\in\mathbb{R}} \frac{t^p}{p!} J(t) dt\right\vert
+ \left\vert\widetilde{R}_j\right\vert 
+ \left\vert\tilde{\text{I}}_j\right\vert\nonumber\\
&\leq \CEZ h_j^{p-1} 
+ h_j^{p-1} \Omega_{nj} \left( \left\vert\int_{t\in\mathbb{R}}  \frac{t^p}{p!} J(t) dt \right\vert  \Vert K\Vert_1^{d-1}
+ \Vert K\Vert_1^{d-1}  \int_{\R} \frac{ \vert t\vert^{p}}{p!} \vert J(t)\vert dt\right)
\end{align*}
with  $\CEZ:=\left\vert \partial_j^p f(w)\int_{t\in\mathbb{R}} \frac{ t^p}{p!} J(t) dt  \right\vert .$

Finally, notice that by continuity of $\partial_j^pf$ (\Cref{Afreg}),  since $\UUn\underset{n\rightarrow\infty}{\longrightarrow}  \lbrace w\rbrace$:
 $$\Omega_{nj}\underset{n\rightarrow\infty}{\longrightarrow} 0.$$

Thus for $n$ large enough:
$$\Omega_{nj} \left( \left\vert\int_{t\in\mathbb{R}}  \frac{t^p}{p!} J(t) dt \right\vert  \Vert K\Vert_1^{d-1}
+ \Vert K\Vert_1^{d-1}  \int_{z_j\in\R} \frac{ \vert z_j\vert^{p}}{p!} \vert J(z_j)\vert dz_j\right)
\leq \frac12 \CEZ,$$
which lead to the result \eqref{e2.lmZbar1} of \Cref{thmRodeo}:
\begin{equation*}
\frac12 \CEZ h_j^{p-1}
\leq \left\vert\E{\bar{Z}_{hj}}\right\vert
\leq \frac32 \CEZ h_j^{p-1}.
\end{equation*}

To obtain the result \eqref{e2.lmZbar2} of \Cref{thmRodeo}, just note that:
\begin{align*}
\E{\vert \bar{Z}_{h1j}\vert}
&=\frac{1}{h_j\prod_{k=1}^d h_k}\int_{u\in\mathbb{R}^d }\left\vert J(\tfrac{w_j-u_j}{h_j})\left(\prod\limits_{k\neq j}{K}(\tfrac{w_k-u_{k}}{h_k})\right)\right\vert f(u)du\\
&= h_j^{-1}\int_{z\in\mathbb{R}^d } \left\vert J(z_j)\left(\prod\limits_{k\neq j}{K}(z_k)\right)\right\vert f(w-Hz)dz\\
 &\leq \CEZA h_j^{-1},
\end{align*}
 with $\CEZA
 :=\Vert f\Vert_{\infty,\UUn} \Vert J\Vert_1 \Vert K\Vert_1^{d-1}$.
 \item We first bound $\bar{Z}_{hij}$ a.s. and its variance.
\begin{align}
\vert \bar{Z}_{hij}\vert
&= \frac{\left\vert J(\frac{w_j-W_{ij}}{h_j})\right\vert\prod\limits_{k\neq j}\left\vert {K}(\tfrac{w_k-W_{ik}}{h_k})\right\vert }{f_X(X_i)h_j\prod_{k=1}^d h_k} 
\nonumber\\
&\leq \frac{\Vert J\Vert_{\infty}\Vert K\Vert_{\infty}^{d-1}}{\delta h_j\prod_{k=1}^dh_k}
= \frac{\CMZbar}{h_j\prod_{k=1}^d h_k}=:\MZbar{hj} \label{eMajZbar as}
\end{align}
 For the variance:
 \begin{align}
 \V{\bar{Z}_{hij}} 
 &\leq \E{\bar{Z}_{hij}^2}
\nonumber\\
 &=\int_{\mathbb{R}^d} J(\tfrac{w_j-u_j}{h_j})^2\left(\prod\limits_{k\neq j}K(\tfrac{w_k-u_k}{h_k})^2\right)\frac{f_{XY}(u)}{f_X(u_{1:d_1})^2h_j^2\prod_{k=1}^dh_k^2} du
\nonumber  \\
 &=\frac{1}{h_j^2 \prod_{k=1}^d h_k } \int_{\mathbb{R}^d} J(z_j)^2\left(\prod\limits_{k\neq j}K(z_k)^2\right) \frac{f(w-Hz)}{f_X(x-(Hz)_{1:d_1})}dz
\nonumber \\
&\leq \frac{\Vert f\Vert_{\infty,\UUn}\Vert J\Vert_2^2\Vert K\Vert_2^{2(d-1)} }{\delta h_j^2 \prod_{k=1}^d h_k }=\frac{\CvZbar}{h_j^2\prod_{k=1}^dh_k}=:\vZbar{hj}.\label{eMajVarZbar}
 \end{align}
 We apply Bernstein's inequality (cf \Cref{lmBernstein}) to $\Zbar{hj}$: 
\begin{align*}
\PP{\BZ{hj}^c}
\leq 2 \exp\left( -\min\left(\frac{n(\tfrac{\seuil{hj}}{2})^2}{4\vZbar{hj}} , \frac{3n\tfrac{\seuil{hj}}{2}}{4\MZbar{hj}}\right)\right)
\end{align*}
Let us compare the rates:
\begin{align*}
&\frac{n(\tfrac{\seuil{hj}}{2})^2}{4\vZbar{hj}} 
\leq \frac{3n\tfrac{\seuil{hj}}{2}}{4\MZbar{hj}}
\\
\Longleftrightarrow\quad  &\Cl\sqrt{\frac{(\log n)^a}{nh_j^2\prod_{k=1}^d h_k}}
=\seuil{hj}
\leq \frac{6\vZbar{hj}}{\MZbar{hj}}
=\frac{6\CvZbar}{\CMZbar h_j}
\\
\Longleftrightarrow\quad  &\prod_{k=1}^d h_k
\geq \frac{\CMZbar^2\Cl^2}{6^2\CvZbar^2} \frac{(\log n)^a}{n}
\\
\Longleftrightarrow\quad  & \text{Cond}_{\bar{Z}}(h).
\end{align*}
So, if $\text{Cond}_{\bar{Z}}(h)$ is satisfied:
\begin{align*}
\PP{\BZ{hj}^c}
\leq 2 e^{\frac{-n(\seuil{hj}/2)^2}{4\vZbar{hj}}}
= 2 e^{\frac{-\delta}{\Vert f\Vert_{\infty, \UUn}}(\log n)^a}= 2 e^{-\gZ}
\end{align*}
\item  We apply Bernstein's inequality (cf \Cref{lmBernstein}) to $\frac1{n}\sum\limits_{i=1}^n \vert \Zbar{hij}\vert $
 using the upper bounds \eqref{eMajZbar as} and \eqref{eMajVarZbar}: 
 \begin{align*}
\PP{\BZA{h}^c}
\leq 2 \exp\left( -\min\left(\frac{n(\CEZA h_j^{-1})^2}{4\vZbar{hj}} , \frac{3n\CEZA h_j^{-1}}{4\MZbar{hj}}\right)\right)
\end{align*}
Let us calculate the rate: by definition of $\CvZbar{hj}$ and $\MZbar{hj}$,
\begin{align*}
\frac{n(\CEZA h_j^{-1})^2}{4\vZbar{hj}}=\frac{\CEZA^2}{4\CvZbar}n\prod\limits_{k=1}^dh_k
\\
\frac{3n\CEZA h_j^{-1}}{4\MZbar{hj}}= \frac{3\CEZA}{4\CMZbar}n\prod\limits_{k=1}^dh_k
\end{align*}
Hence:
 \begin{align*}
\PP{\BA{h}^c}\leq 2 e^{-\CgZA n\prod_{k=1}^d h_k},
\end{align*}
with $\CgZA:= \min\left(\frac{\CEZA^2}{4\CvZbar} ; \frac{3\CEZA}{4\CMZbar}\right)$.
\end{enumerate}
\subsubsection{Proof of \Cref{lmD}}
\begin{enumerate}
\item We decompose $\DZ{hj}$ as follows:
\begin{align*}
\DZ{hj}
:=Z_{hj}-\bar{Z}_{hj}
=\frac{1}{n}\sum\limits_{i=1}^n \left(\frac{f_X(X_i)-\fX(X_i)}{\fX(X_i)}\right) \bar{Z}_{hij}.
\end{align*}
Using $\bar{Z}_{hij}=0$ when $X_i\notin\Ux{h}$:
\begin{align}
\left\vert\DZ{hj}\right\vert
\leq \left\Vert\frac{f_X-\fX}{\fX}\right\Vert_{\infty,\Un} \frac{1}{n}\sum\limits_{i=1}^n \left\vert \bar{Z}_{hij}\right\vert\label{eDecomDZ}.
\end{align}
First we deal with $\left\Vert\frac{f_X-\fX}{\fX}\right\Vert_{\infty,\Un} $.
By definition of $\An$:
\begin{align}
\ind_{\An}\left\Vert\frac{f_X-\fX}{\fX}\right\Vert_{\infty,\Un} 
\leq M_X \left(\frac{(\log n)^d}{n}\right)^{1/2}, \label{eNormeDfX}
\end{align}

Now let us give an upper bound of $\frac{1}{n}\sum\limits_{i=1}^n \left\vert \bar{Z}_{hij}\right\vert$.
Using \Cref{lmZbar}, 
\begin{align*}
\ind_{\BZA{hj}}\frac{1}{n}\sum\limits_{i=1}^n \left\vert \bar{Z}_{hij}\right\vert
&\leq \ind_{\BZA{hj}}\left\vert\frac{1}{n}\sum\limits_{i=1}^n \left\vert \bar{Z}_{hij}\right\vert-\E{\vert \bar{Z}_{h1j}\vert}\right\vert + \E{\vert \bar{Z}_{h1j}\vert}\\
&\leq 2 \CEZA h_j^{-1}
\end{align*}

To conclude, combining this last result with  \eqref{eNormeDfX} and \eqref{eDecomDZ}:
\begin{align*}
\ind_{\BZA{hj}\cap\An}\left\vert\DZ{hj}\right\vert 
&\leq  2 \CEZA M_X h_j^{-1} \left(\frac{(\log n)^d}{n}\right)^{1/2}\\
&\leq  \frac{2 \CEZA M_X}{\Cl (\log n)^{\frac{a}2}} \seuil{hj}= \frac{\CMDZ}{ (\log n)^{\frac{a}2}} \seuil{hj},
\end{align*}
since $\prod\limits_{k=1}^d h_k\leq h_0^d=\frac1{(\log n)^d}$.
\item We decompose $\Delta_{h}$ as follows:
\begin{align*}
\Delta_{h}
:=\f{h}-\fbar{h}
=\frac{1}{n}\sum\limits_{i=1}^n \left(\frac{f_X(X_i)-\fX(X_i)}{\fX(X_i)}\right) \fbar{hi}.
\end{align*}
Using $\fbar{hi}=0$ when $X_i\notin\Ux{h}$:
\begin{align*}
\left\vert\Delta_{h}\right\vert
\leq \left\Vert\frac{f_X-\fX}{\fX}\right\Vert_{\infty,\Un} \frac{1}{n}\sum\limits_{i=1}^n \left\vert \fbar{hi}\right\vert.
\end{align*}
We have proved in  \eqref{eNormeDfX}:
$\ind_{\An}\left\Vert\frac{f_X-\fX}{\fX}\right\Vert_{\infty,\Un} 
\leq M_X\left(\frac{(\log n)^d}{n}\right)^{1/2}$.\\
Let us now give an upper bound of $\frac{1}{n}\sum\limits_{i=1}^n \left\vert \fbar{hi}\right\vert$.
Using \Cref{lmfbar}, 
\begin{align*}
\ind_{\BA{h}}\frac{1}{n}\sum\limits_{i=1}^n \left\vert \fbar{hi}\right\vert
&\leq \ind_{\BA{h}}\left\vert\frac{1}{n}\sum\limits_{i=1}^n \left\vert \fbar{hi}\right\vert-\E{\vert \fbar{hi}\vert}\right\vert + \E{\vert \fbar{h1}\vert}\\
&\leq 2 \CEbar.
\end{align*}
Therefore:
\begin{equation*}
\ind_{\An\cap\BA{h}} \left\vert\Delta_{h}\right\vert
\leq 2\CEbar M_X \left(\frac{(\log n)^d}{n}\right)^{1/2}
\leq  \frac{2\CEbar M_X}{\Cs}(\log n)^{-\frac{a}{2}} \sigma_h,
\end{equation*}
since $\prod\limits_{k=1}^d h_k\leq h_0^d=(\log n)^{-d}$.
\end{enumerate}

\subsubsection{Proof of \Cref{lmKernelBiasUpperBound}}
We first denote 
$$ B:=\int_{\R^{d'}} \left(\prod\limits_{j=1}^{d'} h_j^{-1}K\left(\tfrac{u_j-u'_j}{h_j}\right)\right) f_0(u') du' - f_0(u)\int_{\R^{d'}}   \left(\prod\limits_{j=1}^{d'} K\left(z_j\right)\right)dz.$$
Then we obtain by integration by parts:
\begin{align}\label{eKernelBias}
B:=
\int_{z\in\mathbb{R}^{d'}} \left(\prod\limits_{j=1}^{d'} K(z_j)\right) (f_0(u-h\cdot z)-f_0(u)) dz
\end{align}
For any $z\in\mathbb{R}^{d'}$, we denote $\overline{z}_0:=w$ and  for $k=1:d'$, $\overline{z}_k:=u-\sum\limits_{j=1}^k h_jz_je_j$ (where $\lbrace e_j\rbrace_{j=1}^{d'}$ is the canonical basis of $\R^{d'}$). 
Then, we write:
\begin{align}
f_0(u-h\cdot z)-f_0(u)
=\sum\limits_{k=1}^{d'} f_0(\overline{z}_k)-f_0(\overline{z}_{k-1})
\end{align}
Then we apply Taylor's theorem (cf \Cref{lmTaylor}) to the functions $g_k:t\in[0,1]\mapsto f_0(\overline{z}_{k-1}-th_kz_ke_k)$, $k\in(1:d')$:
\begin{align*}
f_0(\overline{z}_k)-f_0(\overline{z}_{k-1})
=g_k(1)-g_k(0)
=\sum\limits_{l=1}^{p}\frac{(-z_kh_k)^l}{l!}\partial_k^l f_0(\overline{z}_{k-1}) + \rho_k,
\end{align*}
where we denote for short:
\begin{align}\label{edefRhok}
\rho_k:=\rho_k(z,h,u)
=(-h_kz_k)^p \int\limits_{0\leq t_p\leq \dots\leq t_1\leq 1} 
\left(\partial_k^pf_0(\overline{z}_{k-1}-t_ph_kz_ke_k)-\partial_k^pf_0(\overline{z}_{k-1})\right)dt_{1:p}.
\end{align}
We introduce the notation 
$$\text{I}_k:=\int_{z\in\mathbb{R}^{d'}} \left(\prod\limits_{j=1}^{d'} K(z_{j})\right) \rho_k dz$$
 and for any $z\in\mathbb{R}^{d'}$, we denote $z_{-k}\in\R^{d'-1}$ the vector $z$ without its $k^{th}$ variable, then \eqref{eKernelBias} becomes:
\begin{align*}
B
&=
\int_{z\in\R^{d'}} \left(\prod\limits_{j=1}^{d'} K(z_{j})\right) \left(
\sum\limits_{k=1}^{d'} 
\sum\limits_{l=1}^{p}\frac{(-h_k)^l}{l!}  z_k^l  \partial_k^l f_0(\overline{z}_{k-1}) 
 + \rho_k \right)dz
 \\
&=\sum\limits_{k=1}^{d'}\left( \text{I}_k 
+\sum\limits_{l=1}^{p}\frac{(-h_k)^l}{l!}
\int_{z_{-k}\in\R^{d-1}}\partial_k^l f_0(\overline{z}_{k-1}) \left(\prod\limits_{j\neq k} K(z_{j})\right)
\int_{z_k\in\mathbb{R}} z_k^l K(z_k) dz_kdz_{-k} 
\right)
\end{align*}
Since  $K$ has at least $p-1$ zero moments, the terms with $l\leq p-1$ vanish, leading to:
\begin{align}
B
&=\sum\limits_{k=1}^{d'}\left( \text{I}_k
+ \frac{(-h_k)^p \int_{t\in\mathbb{R}} t^p K(t) dt}{p!}
\int_{z_{-k}\in\R^{d'-1}}\partial_k^p f_0(\overline{z}_{k-1}) \left(\prod\limits_{j\neq k} K_j(z_{j})\right)
dz_{-k} \right)
\nonumber\\
&=:\sum\limits_{k=1}^{d'} 
(\text{I}_k 
+\text{II}_k),
\label{eBias2}
\end{align}
with $\text{II}_k:=(-h_k)^p \int_{t\in\mathbb{R}} \frac{t^p}{p!} K(t) dt\int_{z_{-k}\in\R^{d'-1}}\partial_k^p f_0(\overline{z}_{k-1}) \left(\prod\limits_{j\neq k} K(z_{j})\right)
dz_{-k}$.

\section*{Acknowledgement.}
The author is extremely grateful to Claire Lacour and Vincent Rivoirard for suggesting me to
study this problem, for the stimulating discussions, the helpful advices and the careful proofreading.


\bibliographystyle{apalike}
\bibliography{biblio}
%
%
%
%
%
%
%

\end{document}